\documentclass[aip, pof, preprint, floatfix]{revtex4-1}

\usepackage{algorithm}
\usepackage{algpseudocode}
\usepackage{amsmath, amssymb, amsfonts}
\usepackage{bm}
\usepackage{color}
\usepackage{graphicx}
\PassOptionsToPackage{hyphens}{url}\usepackage{hyperref}
\usepackage{siunitx}
\usepackage{subcaption}
\usepackage{tikz}
\usepackage[normalem]{ulem}

\colorlet{RED}{red}
\graphicspath{{../Figures/}}

\newcommand\Rey{\textit{Re}}
\newcommand\Bin{\textit{Bi}}

\newcommand{\sfrac}[2]{\mathchoice
  {\kern0em\raise.5ex\hbox{\the\scriptfont0 #1}\kern-.15em/
   \kern-.15em\lower.25ex\hbox{\the\scriptfont0 #2}}
  {\kern0em\raise.5ex\hbox{\the\scriptfont0 #1}\kern-.15em/
   \kern-.15em\lower.25ex\hbox{\the\scriptfont0 #2}}
  {\kern0em\raise.5ex\hbox{\the\scriptscriptfont0 #1}\kern-.2em/
   \kern-.15em\lower.25ex\hbox{\the\scriptscriptfont0 #2}}
  {#1\!/#2}}

\newcommand{\half}{\frac{1}{2}}

\newcommand*\diff{\mathop{}\!\mathrm{d}}
\newcommand{\mpo}{{m + 1}}

\newcommand{\strainrate}{\dot{\bm{\gamma}}}
\newcommand{\bu}{\bm{u}}
\newcommand{\bx}{\bm{x}}
\newcommand{\bi}{{\bm{i}}}
\newcommand{\bn}{\hat{\bm{n}}}
\newcommand{\stress}{\bm{\tau}}
\newcommand{\norm}[1]{\left\Vert #1 \right\Vert}

\newcommand{\DW}{\left(\frac{\mathcal{D}}{\mathcal{W}}\right)}
\newcommand{\WD}{\left(\frac{\mathcal{W}}{\mathcal{D}}\right)}

\DeclareRobustCommand\sampleline[1]{%
  \tikz\draw[#1] (0,0) (0,\the\dimexpr\fontdimen22\textfont2\relax)
  -- (2em,\the\dimexpr\fontdimen22\textfont2\relax);%
}

\usepackage[colorinlistoftodos]{todonotes}

\begin{document}

\title{An embedded boundary approach for efficient simulations of viscoplastic fluids in three
dimensions}

\author{Knut Sverdrup}
\email{knutsverdrup@gmail.com}
\affiliation{Cavendish Laboratory, University of Cambridge, CB3 0HE, UK}

\author{Ann Almgren}
\affiliation{Lawrence Berkeley National Laboratory, CA 94720, USA}

\author{Nikolaos Nikiforakis}
\affiliation{Cavendish Laboratory, University of Cambridge, CB3 0HE, UK}

\date{\today}

\begin{abstract}
    We present a methodology for simulating three-dimensional flow of incompressible viscoplastic
    fluids modelled by generalised Newtonian rheological equations. It is implemented in a highly
    efficient framework for massively parallelisable computations on block-structured grids. In
    this context, geometric features are handled by the embedded boundary approach, which requires
    specialised treatment only in cells intersecting or adjacent to the boundary. This constitutes
    the first published implementation of an embedded boundary algorithm for simulating flow of
    viscoplastic fluids on structured grids. The underlying algorithm employs a two-stage
    Runge-Kutta method for temporal discretisation, in which viscous terms are treated
    semi-implicitly and projection methods are utilised to enforce the incompressibility
    constraint. We augment the embedded boundary algorithm to deal with the variable apparent
    viscosity of the fluids. Since the viscosity depends strongly on the strain rate tensor,
    special care has been taken to approximate the components of the velocity gradients robustly
    near boundary cells, both for viscous wall fluxes in cut cells and for updates of apparent
    viscosity in cells adjacent to them. After performing convergence analysis and validating the
    code against standard test cases, we present the first ever fully three-dimensional simulations
    of creeping flow of Bingham plastics around translating objects. Our results shed new light on
    the flow fields around these objects.
\end{abstract}


\maketitle 




\section{Introduction}
\label{sec:introduction}

Although a lot of effort has been made towards simulations of viscoplastic fluids, their numerical
treatment is by nature much more expensive than that of fluids which do not exhibit a yield stress
\cite{saramito2017progress, mitsoulis2017numerical}. Consequently, researchers interested in these
flows are often limited by the computational cost associated with simulating them, resulting in the
majority of such work being restricted to two-dimensional (2D) and/or steady-state problems. This is
especially true for fluid-structure interaction problems with many particles, such as the recent
work by Koblitz et al.~\cite{koblitz2018direct}, which would not be practically possible within
current software packages without considering the 2D steady-state problem. Having said that, there
are notable advances being made, such as a study on the transition to turbulence of 
viscoplastic fluids past a cylinder in three dimensions (3D), which was simulated using
Papanastasiou regularisation for small Bingham numbers \cite{kanaris2015transition}. A recent
example of unsteady flow with fluid particle interactions is the study of time-dependent
hydrodynamic interaction of 2D particles by Chaparian et al.~\cite{chaparian2018inline} 

Numerical simulations in the presence of complex boundaries can broadly be sorted by whether they
use structured or unstructured grids. Both types of grids are widely utilised, and an overview of
the usage of each in published simulations of yield stress fluids is provided in the review by
Saramito and Wachs \cite{saramito2017progress}. By definition, the difference between structured
and unstructured grids is that the former requires regular connectivity of cells, while the latter
does not. Consequently, the two types of grids have different advantages. Unstructured meshes have
the advantage of greater geometric generality, while structured grids, which can incorporate
geometric complexity with an embedded boundary approach, require storage of less geometric
information. Algorithms on structured meshes can therefore exploit the excellent data storage for
higher computational efficiency. 

Non-Cartesian geometries are cut out of the underlying grid by storing local data representing the
interface within each cell which contains part of the geometry. The methodology has several
attractive qualities: rapid mesh generation regardless of complex geometries
\cite{roosing2019fast}; avoidance of locally skewed grids; inherent compatibility with quad- and
octree adaptive mesh refinement (AMR); and retainment of the efficient and user-friendly data
storage associated with structured grids. Special consideration must be taken to augment
computational stencils and other algorithmic tools for and near the cut cells, but the technique
nevertheless offers a rapid and relatively simple way of incorporating non-rectangular geometries
without sacrificing the efficiency associated with structured grids.

Over the last decades, the embedded boundary method has matured and become a robust tool. The idea
was first used by Purvis and Burkhalter in 1979 \cite{purvis1979prediction}, and subsequently,
Wedan and South \cite{wedan1983method} to solve potential flow problems. Throughout the 1980s, the
method was extended to solve the compressible Euler equations \cite{clarke1986euler,
falle1988adaptive, berger1989adaptive}, which forms the basis for the general hyperbolic treatment
of Colella et al.~which we follow \cite{colella2006cartesian}. Since we are studying an
incompressible system, we also require the solution of Poisson equations, which was published by
Johansen and Colella in 1998 \cite{johansen1998cartesian}; a feat which they and McCorquodale
naturally extended to solutions of the heat equation a few years later
\cite{mccorquodale2001cartesian}. Developing novel and improved schemes for embedded boundaries is
still an active area of research \cite{berger2017cut, gokhale2018dimensionallyHCL,
gokhale2018dimensionallyCNS}.

In a recent publication \cite{sverdrup2018highly}, we presented high-performance software capable
of simulating 3D flow of viscoplastic fluids in time, utilising structured
AMR. There, the efficacy of highly parallelisable structured meshing was
demonstrated, though computational domains were restricted to rectangular and right rectangular
prisms in two and three dimensions, respectively. In this work, we extend the software package to
non-rectangular geometries through the use of EBs. This was a natural extension
of our work, and the EB approach was particularly well-suited to the structured mesh
discretisation. The technique has not been utilised for viscoplastic fluid flow previously, but has
been expected to perform quite well \cite{wachs2019computational}.

This paper constitutes an entirely novel utilisation of EB techniques to treat flow problems
involving generalised Newtonian fluids, and more specifically yield-stress fluids. While evaluating
the software, the extension has allowed us to obtain rich insight into the flow fields around
objects moving through Bingham fluids. In particular, the three-dimensional effects on the yield
surface of a sphere in such a configuration are properly investigated for the first time, and we
show how more general, asymmetric flows can be simulated just as easily.

In section \ref{sec:mathematical}, we will formulate the mathematical description of the system of
governing partial differential equations and discuss relevant fluid rheologies. Section
\ref{sec:numerical} is devoted to the numerical algorithm employed to simulate the fluid flow,
including treatment of embedded boundaries. Thorough validation is performed in section
\ref{sec:verification}, before we evaluate the code for more demanding problems with genuinely
three-dimensional effects in section \ref{sec:evaluation}. Section \ref{sec:conclusions} concludes
the article.


\section{Mathematical formulation}
\label{sec:mathematical}

Our domain $\Omega \subset \mathbb{R}^3$ has a boundary denoted by $\partial \Omega$. We take the
gradient of a vector $\bu$ as the tensor with components
\begin{equation}
    (\nabla \bu)_{ij} = \frac{\partial u_j}{\partial x_i},
\end{equation}
while the divergence of a rank-2 tensor field $\stress$ is defined such that
\begin{equation}
    (\nabla \cdot \stress)_j = \sum_{i=1}^{3} \frac{\partial \tau_{ij}}{\partial x_i}.
\end{equation}

Variables are functions of position $\bx \in \Omega$ and time $t \geq 0$. We denote by $\rho \in
\mathbb{R}$ the material density. The velocity field is introduced as $\bu (\bx, t) \in
\mathbb{R}^3$, with components $u$, $v$ and $w$. The Cauchy stress tensor $\bm{\sigma} (\bx, t)$ is
defined as the sum of isotropic and deviatoric parts, $\bm{\sigma} = -p \bm{I} + \stress$. Here,
the pressure $p (\bx, t) \in \mathbb{R}$ is multiplied by the identity tensor, while the deviatoric
part of the stress tensor is denoted $\stress(\bx,t) \in \mathbb{R}^{3 \times 3}_{\rm sym}$.

\subsection{Governing partial differential equations}

We consider time-dependent flow of incompressible, generalised Newtonian fluids. Denoting external
body forces by $\bm{f}$, the relevant governing equations are
\begin{subequations}
    \label{eq:governing}
    \begin{align}
        \label{eq:governing:cauchy}
        \frac{\partial \bu}{\partial t} + \bu \cdot \nabla \bu
        &= \frac{1}{\rho} \left( -\nabla p + \nabla \cdot \stress + \bm{f} \right)
        & {\rm in} \quad \Omega \\
        \label{eq:governing:incompressibility}
        \nabla \cdot \bu &= 0
        & {\rm in} \quad \Omega \\
        \label{eq:governing:stress}
        \stress &= \eta \left( \norm{\strainrate} \right) \strainrate
        & {\rm in} \quad \Omega \\
        \label{eq:governing:no-slip}
        \bu &= \bu_{\rm BC}
        & {\rm on} \quad \partial \Omega
    \end{align}
\end{subequations}
where the strain rate tensor $\strainrate = \nabla \bu + \left( \nabla \bu \right)^\top$ is (twice)
the symmetric part of the velocity gradient, and $\norm{\cdot}$ denotes the scaled Frobenius norm
such that $\norm{\strainrate} = \sqrt{\half {\rm tr} \left( \strainrate \strainrate^\top \right)}$,
as is customary for convenience in viscoplastic fluid mechanics. Equation
\eqref{eq:governing:cauchy} is Cauchy's momentum balance, \eqref{eq:governing:incompressibility} is
the incompressibility constraint, and \eqref{eq:governing:stress} relates the stress to the
specific rheological equation for a generalised Newtonian fluid model through the apparent
viscosity $\eta \left( \norm{\strainrate} \right)$.  The perfect no-slip boundary condition which
we employ in \eqref{eq:governing:no-slip} is the simplest and most common way to deal with wall
effects for viscous flow, although many cases require more advanced treatments of the solid-fluid
interactions, such as the stick-slip condition for Bingham fluids \cite{muravleva2017axisymmetric,
muravleva2018squeeze}. 

\subsection{Rheology}

An applied shear strain acting on a fluid causes flow through viscous deformation, and rheological
equations give the relationship between the stress and strain tensors, their temporal derivatives
and physical variables such as time, temperature and pressure \cite{barnes1989introduction}. As
seen from \eqref{eq:governing:stress}, we restrict ourselves to relatively simple equations of
state on the form $\stress = \stress(\strainrate)$. In other words, the stress response is solely
dependent on the strain rate tensor. Furthermore, the relation is characterised by an apparent
viscosity $\eta$, which is a function of the second invariant of the strain rate tensor, ${\rm tr}
\left( \strainrate \strainrate^\top \right)$. For Newtonian fluids, a constant viscosity
coefficient quantifies the proportionality between stress and strain rate. Due to the analogy to
this simplest of viscous models, the fluids which we consider are called generalised Newtonian
fluids.

Although generalised Newtonian fluids cannot capture exotic rheological phenomena such as
thixotropy and rheopecty \cite{mewis1979thixotropy, barnes1997thixotropy}, they can still capture a
rich variety of fluid dynamics. Physical phenomena which are widely encountered in applications
include shear induced thinning and thickening, and the existence of a yield stress, below which
flow does not occur. In table \ref{tab:rheology}, we list the fluids models which have been
implemented in our software package, and relevant references. Newtonian fluids have a constant
dynamic viscosity coefficient $\mu$, while shear thinning and thickening is modelled through the
power law model of Ostwald and de Waele, $\eta = \kappa \norm{\strainrate}^{n-1}$,
in which $\mu$ is replaced by the consistency $\kappa$ and a flow index $n$ is introduced. Its
value is between zero and one for shear thinning fluids (pseudoplastics) and larger than one for
shear thickening fluids (dilatants). For $n=1$, we recover the Newtonian model.

Viscoplastics, commonly referred to as yield stress fluids, are characterised by a yield stress
$\tau_0$, which is a threshold value that must be overcome by the applied shear stress for any flow
to occur at all. Since viscoplasticity is the most interesting feature in our fluid models, and at
the same time the most challenging to capture numerically, we will be using the simplest such model
to validate and showcase our software. It is the Bingham plastic fluid, which has a linear
relationship between stress and strain rate above the yield limit:
\begin{equation}
    \begin{cases}
        \norm{\stress} \leq \tau_0 \,,
        & {\rm if } \quad \norm{\strainrate} = 0 \\
        \stress = \left( \mu + \frac{\tau_0}{\norm{\strainrate}} \right) \strainrate \,,
        & {\rm if } \quad \norm{\strainrate} > 0
    \end{cases} .
    \label{eq:bingham:stress}
\end{equation}
The Bingham plastic rheological model thus separates the flow into two separate states. We refer to
the case of zero strain rate as an unyielded state, while viscous flow occurs for the yielded
state, when $\norm{\stress} > \tau_0$. For the yielded Bingham fluid, we can rewrite
\eqref{eq:bingham:stress} as a generalised Newtonian fluid, taking the apparent viscosity function
as
\begin{equation}
    \eta = \mu + \frac{\tau_0}{\norm{\strainrate}} .
    \label{eq:bingham:viscosity}
\end{equation}
Since this results in a singularity in the limit of zero strain rate, we utilise the regularisation
approach introduced by Papanastasiou \cite{papanastasiou1987flows}. By introducing a small
regularisation parameter $\varepsilon$ and multiplying the singular term by one minus a decaying
exponential, we remove the singularity. Consequently, we limit the maximum possible viscosity in
the fluid and obtain an apparent viscosity function defined for all values of the rate-of-strain
tensor,
\begin{equation}
    \eta = \mu + \frac{\tau_0}{\norm{\strainrate}} 
    \left( 1 - e^{-\norm{\strainrate} / \varepsilon} \right) .
    \label{eq:papabing:viscosity}
\end{equation}
Although the regularisation approach removes the singularity and, consequently, enables us to
simulate the viscoplastic behaviour with relative ease, it is important to note that we are
actually solving a slightly different problem as long as $\varepsilon > 0$. In order to solve the
unregularised problem, it is necessary to use augmented Lagrangian methods or similar algorithms
from non-convex optimisation, which are computationally expensive by comparison. Papanastasiou regularisation
is extensively used in state of the art fluid mechanics codes and contemporary research, and
although the debate of its pros and cons compared to the single-parameter Bingham model is
interesting, we will not revisit it here. Instead, we direct the interested reader to the
discussions by Frigaard and Nouar \cite{frigaard2005usage} and Dinkgreve et
al.~\cite{dinkgreve2017everything}, in addition to some recent review papers
\cite{balmforth2014yielding, saramito2017progress, mitsoulis2017numerical}.

The Herschel-Bulkley model generalises power-law fluids to the realm of viscoplasticity in the same
way as the Bingham plastic generalises Newtonians. Finally, we include a model by de Souza Mendes
and Dutra \cite{mendes2004viscosity}, which has the same flexibility and desirable attributes as the
regularised Herschel-Bulkley model, but which seeks to account for the limiting behaviour at small
strain rate as a physical phenomena, rather than as a convenient numerical tool. For a further
discussion on viscoplastic fluids and their numerical treatment, we refer to our previous paper
\cite{sverdrup2018highly}.

\begin{table}
    \centering
    \begin{tabular}{lc}
        \toprule
        Fluid model &
        $\eta \left( \norm{\strainrate} \right)$ \\
        \colrule
        Newtonian &
        $\mu$ \\
        Power-law \cite{de1925aenderung} &
        $\kappa \norm{\strainrate}^{n-1}$ \\
        Bingham \cite{bingham1916investigation, oldroyd1947rational, papanastasiou1987flows} &
        $\mu + \frac{\tau_0}{\norm{\strainrate}}
        \left( 1 - e^{-\norm{\strainrate}/\varepsilon} \right)$ \\
        Herschel-Bulkley \cite{herschel1926konsistenzmessungen, papanastasiou1987flows} &
        $\kappa \norm{\strainrate}^{n-1} + \frac{\tau_0}{\norm{\strainrate}}
        \left( 1 - e^{-\norm{\strainrate}/\varepsilon} \right)$ \\
        Souza Mendes-Dutra  \cite{mendes2004viscosity} &
        $\kappa \norm{\strainrate}^{n-1} + \frac{\tau_0}{\norm{\strainrate}}
        \left( 1 - e^{-\eta_0 \norm{\strainrate}/\tau_0} \right)$ \\
        \botrule
    \end{tabular}
    \caption{Generalised Newtonian fluid models available in the fluid solver.}
    \label{tab:rheology}
\end{table}


\section{Numerical algorithm}
\label{sec:numerical}

This section is devoted to the details of the numerical algorithm utilised in order to solve
\eqref{eq:governing} efficiently, taking advantage of modern supercomputer architectures. The basis
for the fluid solver is a projection method for solving the variable density incompressible
Navier-Stokes equations on an adaptive mesh hierarchy, as described for the constant viscosity case
by Almgren et al.~ \cite{almgren1998conservative}, with extensions to low Mach number reacting
flows with temperature-dependent viscosity provided by Pember et al.~\cite{pember1998adaptive} and
Day and Bell \cite{day2000numerical}, among others. By using the software framework AMReX, the
resulting implementation is such that the code can be run on architectures from single-core laptops
through to large-scale distributed cluster computers.

AMReX \cite{zhang2019amrex} is a mature, open-source software framework for
building massively parallel structured AMR applications. It contains extensive software support
for explicit and implicit grid-based operations. Multigrid solvers, including those for tensor
systems, are included for cell-based and node-based data. A hybrid MPI/OpenMP approach is used 
for parallelisation; in this model individual grids are distributed to MPI ranks, and OpenMP is
used to thread over logical tiles within the grids. Applications based on AMReX have demonstrated
excellent strong and weak scaling up to hundreds of thousands of cores \cite{almgren2010castro,
almgren2013nyx, dubey2014survey, habib2016hacc}. For a demonstration of scalability in the specific
context of yield stress fluids, we refer to our previous paper on the subject
\cite{sverdrup2018highly}.

The functionality provided by the AMReX framework facilitates simple solutions to many of the
complex subproblems we encounter. For example, all linear systems solved in our software utilise
built-in multilevel geometric multigrid solvers \cite{wesseling2001geometric} for EB systems, with
the biconjugate gradient stabilised method (BiCGSTAB) \cite{van1992bi} or other Krylov solvers
available for use at the coarsest multigrid level. For the sake of brevity, we omit a full,
low-level description of the algorithmic implementation and point the interested reader to the code
repository \footnote{\url{https://github.com/AMReX-Codes/incflo}}, and the documentation for AMReX
\cite{zhang2019amrex}.

\subsection{Incompressible flow solver}

For clarity, we omit treatment of boundaries and communication across refinement levels
in the following description of the incompressible flow solver. Instead, we focus on the steps
required to move the system forward one time step on a single refinement level. Although the
software is general enough to handle variable-density incompressible flows through a conservative
advection of $\rho$, we only consider cases where it is constant, and consequently do not discuss
conservative density updates, which in any event are trivial compared to the velocity terms. Note
also that there are no external forces for the cases in this paper, so $\bm{f} = 0$ throughout. 

\subsubsection{Time step constraint}

It is essential to use a time step size which provides numerical stability and accurate results for
the temporal advancement scheme, but we would like to use the largest permissible $\Delta t$.  As a
foundation, we take the thorough derivation by Kang et al.~\cite{kang2000boundary} which arrives at
a time step given by 
\begin{equation}
    \Delta t = \frac{C_{CFL}}{\half \left( C_C + C_V + \sqrt{(C_C + C_V)^2 + 4 C_F^2} \right)} .
    \label{eq:cfl:dt}
\end{equation}
Here, $C_{CFL} \leq 1$ is a parameter called the CFL coefficient. It is set to 0.5 by default, but
can be reduced to smaller values for challenging flow problems. The other coefficients, $C_C$,
$C_V$ and $C_F$, are constraint coefficients due to convective, viscous and forcing terms,
respectively. The convective coefficient is given by 
\begin{equation}
    C_C = \max_{\Omega} 
    \left( \frac{|u|}{\Delta x} + \frac{|v|}{\Delta y} + \frac{|w|}{\Delta z} \right) ,
    \label{eq:cfl:convective}
\end{equation}
and the coefficient due to external forcing is
\begin{equation}
    C_F = \sqrt{\frac{|f_x|}{\Delta x} + \frac{|f_y|}{\Delta y} + \frac{|f_z|}{\Delta z}} .
    \label{eq:cfl:forcing}
\end{equation}
For the viscous coefficient, Kang et al.~\cite{kang2000boundary} utilised
\begin{equation}
    C_{V, \, \rm Kang} = 2 \max_{\Omega} \left( \frac{\mu}{\rho} \left( \frac{1}{\Delta x^2} +
    \frac{1}{\Delta y^2} + \frac{1}{\Delta z^2} \right) \right) ,
    \label{eq:cfl:viscous}
\end{equation}
but that work was restricted to Newtonian flows. Direct substitution of $\mu$ by the apparent Bingham
viscosity $\eta$ in \eqref{eq:cfl:viscous}, results in a computed $\Delta t$ which is overly
restrictive for viscoplastic flows where the Bingham number is large.  Syrakos et
al.~\cite{syrakos2016cessation} demonstrated that the time step size for such flows is actually
dependent on the ratio of the Reynolds to Bingham numbers. In their article, they found that it was
sufficient to utilise a time step 
\begin{equation}
    \Delta t_{\rm Syrakos} \propto \frac{\Rey}{\Bin + 1} \Delta x^2 ,
    \label{eq:dt:syrakos}
\end{equation}
where the proportionality factor is $\mathcal{O}(1)$. Notably, \eqref{eq:dt:syrakos} does not need
to take the level of regularisation into account (they used $\varepsilon = 1 / 400$). In
fact, inserting the numbers from their study into \eqref{eq:cfl:viscous} gives the ratio between
the two as 
\begin{equation}
    \frac{\Bin + 1}{\Bin / \varepsilon + 1} \approx \varepsilon ,
    \label{eq:dt:ratio}
\end{equation}
so that the time step is much smaller than it needs to be, by a factor inversely proportional to
the regularisation parameter. We therefore remove the dependency of \eqref{eq:cfl:viscous} on the
regularisation parameter, and instead replace it by 
\begin{equation}
    C_V = \frac{2 \mu}{\rho} (\Bin + 1) \max_{\Omega} 
    \left( \frac{1}{\Delta x^2} + \frac{1}{\Delta y^2} + \frac{1}{\Delta z^2} \right) .
    \label{eq:cfl:viscous:new}
\end{equation}

\subsubsection{Temporal integration}

We follow a method-of-lines (MOL) approach, in which the momentum equation as given by
\eqref{eq:governing:cauchy} is only discretised in time, so that Runge-Kutta type schemes for
ordinary differential equations (ODEs) can be utilised for the temporal advancement. We affix a
superscript to the simulated variables at the $m$-th time step, so that the time itself is $t^m$
and the velocity profile $\bu^m$. Algorithm \ref{alg:advance} gives the pseudocode to advance the
simulation from time step $m$ to $m+1$.

\begin{algorithm}[H]
    \caption{Advance simulation one time step}
    \label{alg:advance}
    \begin{algorithmic}[1]
        \Function{Advance}{$t^m, \bu^m, p^m$}
            \State \Call{ApplyBoundaryConditions}{$\partial \Omega$}
            \State $\Delta t := $ \Call{TimeStepSize($t^m, \bu^m, p^m$)}{}
            \Procedure{Predictor}{}
                \State $\bm{C}^m, \eta^m, \bm{V}^m := $ \Call{DerivedQuantities}{$\bu^m$}
                \Statex \Comment{Defined in algorithm \ref{alg:derived}}
                \State $\bm{C} \leftarrow \bm{C}^m$
                \State $\eta \leftarrow \eta^m$
                \State $\bm{V} \leftarrow \bm{V}^m$
                \State $\bm{\psi} \leftarrow \rho \bu^m
                        + \Delta t \left( \rho \bm{C} + \bm{V} - \nabla p^m + \bm{f} \right)$
                \State $\tilde{\bu}^* := $
                        \Call{UpdateVelocity}{$\Delta t, \eta, \bm{\psi}$}
                \State $\bu^*, p^* := $ \Call{NodalProjection}{$\Delta t, \tilde{\bu}^*$}
            \EndProcedure
            \Procedure{Corrector}{}
                \State $\bm{C}^*, \eta^*, \bm{V}^* := $ \Call{DerivedQuantities}{$\bu^*$}
                \Statex \Comment{Defined in algorithm \ref{alg:derived}}
                \State $\bm{C} \leftarrow \half \left( \bm{C}^m + \bm{C}^* \right)$
                \State $\eta \leftarrow \half \left( \eta^m + \eta^* \right)$
                \State $\bm{V} \leftarrow \half \left( \bm{V}^m + \bm{V}^* \right)$
                \State $\bm{\psi} \leftarrow \rho \bu^m
                        + \Delta t \left( \rho \bm{C} + \bm{V} - \nabla p^* + \bm{f} \right)$
                \State $\tilde{\bu}^\mpo := $
                        \Call{UpdateVelocity}{$\Delta t, \eta, \bm{\psi}$}
                \State $\bu^\mpo, p^\mpo := $ \Call{NodalProjection}{$\Delta t, \tilde{\bu}^\mpo$}
            \EndProcedure
            \State $t^\mpo := t^m + \Delta t$
            \State $m \leftarrow m + 1$
        \EndFunction
    \end{algorithmic}
\end{algorithm}

After applying boundary conditions and computing the time step size, we perform a two-step
Runge-Kutta scheme with a predictor and a corrector step. The pseudocode in each of the procedures
\textsc{Predictor} and \textsc{Corrector} has been written so as to highlight the similarities
between them. Although we are dealing with a multi-step algorithm including projections within each
procedure, this elucidates the classic Runge-Kutta scheme underlying the temporal integration.

Before going into the specifics of the algorithm substeps, we need to mention the spatial
discretisation of our physical variables. Velocities and derived quantities are all computed at
cell centres, while the pressure is computed at nodes. Consequently, pressure gradients are stored
at cell centres, and can be added directly in the momentum balance.

\subsubsection{Computing derived quantities}

The first step within each of the procedures is to compute the quantities which are directly
derived from the velocity field and its gradients. The relevant pseudocode is given in algorithm
\ref{alg:derived}. In order to accurately capture convection within the fluid, a slope-based
upwinding procedure is utilised. Velocity slopes are computed within each cell by comparing values
in neighbouring cells, according to the algorithm due to Colella \cite{colella1985direct}. These
slopes are then used to extrapolate the input velocity to cell faces. With one extrapolated value
from each of the two cell adjacent to the face, a simple upwinding algorithm is utilised. The
normal velocity components are treated first, and subsequently used as the background field when
upwinding the transverse components.

The upwinded, extrapolated values at faces allow us to compute the convective term ${\bm{C} = - \bu
\cdot \nabla \bu^F}$.  Before doing this, however, we enforce the incompressibility constraint in
\eqref{eq:governing:incompressibility} by projecting the face-based extrapolated velocity
$\tilde{\bu}^F$ onto the space of solenoidal vector fields.  This is necessary since the values
extrapolated by upwinding the slopes are not necessarily divergence-free. The Helmholtz
decomposition allows us to write any vector field as a sum of solenoidal and irrotational parts, so
we define
\begin{equation}
    \label{eq:helmholtz}
    \tilde{\bu}^F = \bu^F + \nabla \phi ,
\end{equation}
where $\bu^F$ has zero divergence and $\nabla \phi$, being a scalar gradient, is irrotational.
Taking the divergence of \eqref{eq:helmholtz} leads to a Poisson equation for $\phi$,
\begin{equation}
    \nabla^2 \phi = \nabla \cdot \tilde{\bu}^F ,
\end{equation}
which is straightforward to solve and allows us to compute the divergence-free velocity field
$\bu^F$. Since the three components of $\bu$ are extrapolated to separate faces, the unknown scalar
$\phi$ must be cell-centred in order for its gradients to end up on the corresponding faces.  For
this reason, we refer to this projection as the cell-centred projection, although it has previously
been referred to as the MAC projection due to its historical links to the spatial discretisation in
the original marker and cell method \cite{harlow1965numerical}. 

\begin{algorithm}[H]
    \caption{Compute derived quantities: convective term, apparent viscosity and explicit viscous
    term}
    \label{alg:derived}
    \begin{algorithmic}[1]
        \Function{DerivedQuantities}{$\bu$}
            \State $s = $\Call{ComputeSlopes}{$\bu$}
            \State $\tilde{\bu}^F = $ \Call{ExtrapolateToFaces}{$\bu, s$}
            \State $\bu^F = $ \Call{CellCentredProjection}{$\tilde{\bu}^F$}
            \State $\bm{C} = - \bu \cdot \nabla \bu^F$
                \Comment{Convective term}
                \State $\norm{\strainrate} 
                = \norm{\nabla \bu^F + \left( \nabla \bu^F \right)^\top}$
                \Comment{Strain rate magnitude}
            \State $\eta = \eta(\norm{\strainrate})$
                \Comment{Apparent viscosity}
            \State $\bm{V} = \frac{1}{\rho} \nabla \cdot
                    \left( \eta \left( \nabla \bu \right)^\top \right)$
                \Comment{Non-linear stress divergence}
            \State \textbf{return} $\bm{C}, \eta, \bm{V}$
        \EndFunction
    \end{algorithmic}
\end{algorithm}

With the convective terms in place, the next step is to compute the viscous terms. Since viscous
terms dominate in many of the problems we are interested in, and since they can lead to overly
restrictive time step criteria when treated explicitly, we opt for a semi-implicit temporal
discretisation. Noting that the stress divergence is
\begin{equation}
    \label{eq:divtau}
    \nabla \cdot \stress
    = \nabla \cdot \left( \eta \strainrate \right)
    = \nabla \cdot \left( \eta \nabla \bu \right)
    + \nabla \cdot \left( \eta \left( \nabla \bu \right)^\top \right) ,
\end{equation}
and thus contains one term with purely unmixed derivatives in $\bu$, and one with mixed and
transverse terms, we treat the former implicitly and the latter explicitly. This is due to the fact
that the unmixed result is of a compatible layout with the temporal derivative of $\bu$, and can
readily be set up for a linear solve. The explicit viscous term, derived from the input velocity
is therefore
\begin{equation}
    \label{eq:explicit-viscous}
    \bm{V} = \frac{1}{\rho} \nabla \cdot \left( \eta \left( \nabla \bu \right)^\top \right) ,
\end{equation}
where $\eta$ (via $\norm{\strainrate}$) is calculated from the face-centred extrapolated velocity
components $\bu^F$, as shown on lines 6-8 in algorithm \ref{alg:derived}. Note that one of the
rheological equations in table \ref{tab:rheology} must be specified for the simulation.

In the predictor, the derived quantities from the input velocity $\bu^m$ are used directly in the
velocity update. In the corrector, on the other hand, the derived quantities are first computed
from the predictor's output velocity $\bu^*$, before the average is calculated and utilised in the
update. See lines 6-8 and 15-17 in algorithm \ref{alg:advance}. The semi-implicit velocity update
consists of solving the system
\begin{equation}
    \label{eq:implicit-viscous}
    \left( \rho - \Delta t \nabla \cdot \left( \eta \nabla \right) \right) \tilde{\bu}
    = \rho \bu + \Delta t \left( \rho \bm{C} + \bm{V} - \nabla p + \bm{f} \right)
\end{equation}
with respect to $\tilde{\bu}$, the new-time velocity.

\subsubsection{Incompressibility constraint}

It is necessary to apply another projection in order to enforce the incompressibility constraint
for the new velocity. Conveniently, we can update the pressure at the same time. The two equations
\begin{subequations}
\begin{align}
    \frac{\tilde{\bu} - \bu^{\rm old}}{\Delta t}
    &= - \bu \cdot \nabla \bu + \frac{1}{\rho} \left( \nabla \cdot \stress + \bm{f} \right) \\
    \frac{\bu^{\rm new} - \tilde{\bu}}{\Delta t}
    &= - \frac{1}{\rho} \nabla p
\end{align}
\end{subequations}
sum to a MOL discretised version of \eqref{eq:governing:cauchy}, and the latter is of the form of a
Helmholtz decomposition, just like \eqref{eq:helmholtz}. We therefore add the pressure gradient
term back to the new-time velocity,
\begin{equation}
    \tilde{\bu} \leftarrow \tilde{\bu} + \frac{\Delta t}{\rho} \nabla p
\end{equation}
before solving the Poisson equation
\begin{equation}
    \nabla \cdot \left( \frac{\Delta t}{\rho} \nabla p^{\rm new} \right)
    = \nabla \cdot \tilde{\bu} ,
\end{equation}
for $p^{\rm new}$, and obtain the new, divergence-free velocity field
\begin{equation}
    \bu^{\rm new} = \tilde{\bu} - \frac{\Delta t}{\rho} \nabla p^{\rm new} .
\end{equation}
Note that in this projection, all velocity components are stored on cell centres, and the pressure
is thus nodal. Consequently, we refer to it as the nodal projection. It is in fact a
second-order accurate approximate projection method, the likes of which have been thoroughly
analysed by Almgren et al.~\cite{almgren2000approximate}. The predictor outputs the velocity field
$\bu^*$ and pressure $p^*$, while for the corrector the corresponding variables are $\bu^\mpo$ and
$p^\mpo$. 

After the corrector, $\Delta t$ is added to the current time, and $m$ is incremented, before
continuing on to the next time step.

\subsection{Embedded boundaries}

The EB approach allows us to retain the structured adaptive mesh which AMReX is built for,
while simulating flow in non-rectangular domain boundaries. An arbitrary implicit signed distance
function is used to describe the geometry, which is then discretised as planar intersections with
each cell. The intersections are continuous at cell faces, which means that they are piecewise
linear everywhere. Cells are identified by their index vector $\bi = (i,j,k)$.

\begin{figure*}
    \centering
    \begin{subfigure}{0.4\textwidth}
        \includegraphics[width=\columnwidth]{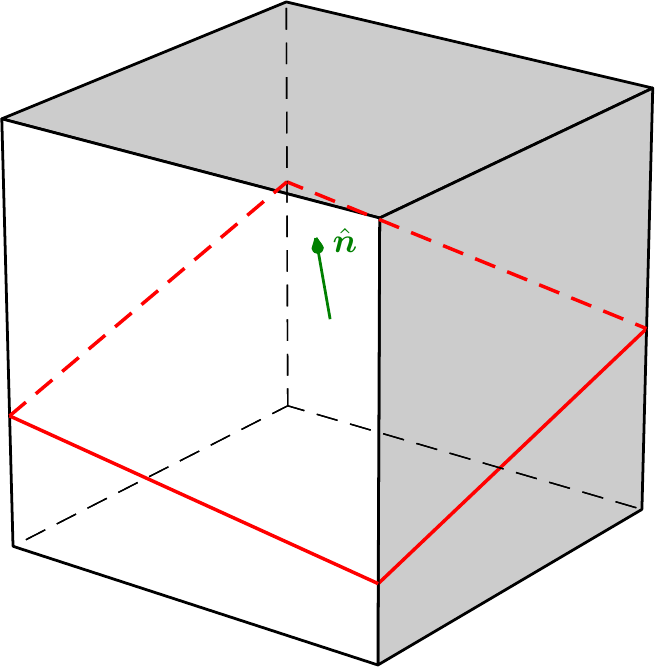}
        \label{fig:cut_cell_3d}
        \caption{3D view}
    \end{subfigure}
    \hspace{2em}
    \begin{subfigure}{0.5\textwidth}
        \includegraphics[width=\columnwidth]{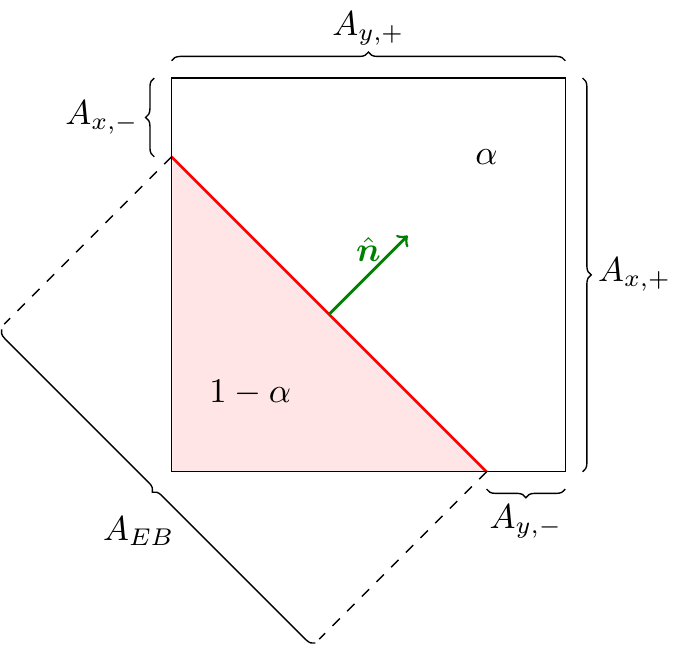}
        \label{fig:cut_cell_2d}
        \caption{2D view of $xy$-plane}
    \end{subfigure}
    \caption
    {
        Schematic illustration of the Cartesian cut cell viewed (a) in 3D and (b) through a slice
        with constant $z$. The plane representing the embedded boundary is marked in red,
        and its surface normal vector $\bn$ points from the covered solid region to the uncovered
        fluid domain with volume fraction $\alpha$. The right plot also shows the relevant surface
        areas. 
    }
    \label{fig:cut_cell}
\end{figure*}

We store a flag in each cell marking it as either uncovered (normal), covered (ignored), or cut. In
the latter case, additional data is stored within the cell, so that we can take geometrical
information into account for computations in that cell. Since we only consider single-valued cut
cells, only four numbers are necessary to uniquely define the cut cell, namely the three components
of the boundary surface unit normal $\bn_{EB}$ plus the volume fraction $\alpha \in (0,1)$ of fluid
within the cell. We orient $\bn_{EB}$ so that it points into the fluid domain, and let $\alpha = 0$
for covered cells and $\alpha = 1$ for uncovered ones. Although these quantities describe the cut
cell, several helpful additions are available in the EB framework provided by AMReX, such as the
uncovered area fractions on each face, the volume centroid of the fluid, and the
area and centroid of the embedded boundary. Figure \ref{fig:cut_cell} illustrates the cut cell and
the relevant quantities for our computations. For details on the EB implementation, we refer to the
AMReX documentation, and for an explanation of how EB data is computed from a signed distance
function, the reader is referred to the paper by Gokhale et al.~\cite{gokhale2018dimensionallyCNS}

All variables in our simulations are stored at the same cell positions, regardless of the cell
type. This has the strong advantage of allowing us to avoid altering computational
stencils for calculating extrapolated values on faces or approximating gradients based on
neighbouring cells which are cut. We only need to worry about the cut cells themselves, which can
have covered (and thus undefined) neighbours. The EB tools built in to the linear solvers for
systems of linear equations (solutions to the Poisson equations) are based on the work of Johansen
and Colella \cite{johansen1998cartesian}.

\subsubsection{Flux computations}

Our main challenge algorithmically (apart from the linear solves, for which AMReX has built-in EB
support) is to successfully compute the derived quantities within each cut cell, i.e.~the convective
term $\bm{C}$, the rate-of-strain tensor magnitude $\norm{\strainrate}$, the apparent viscosity
$\eta$ and the viscous term $\bm{V}$. It is crucial that this is done in a manner which avoids
time step constraints resulting from small cut cell volumes.

\begin{figure}
    \centering
    \includegraphics[width=0.8\columnwidth]{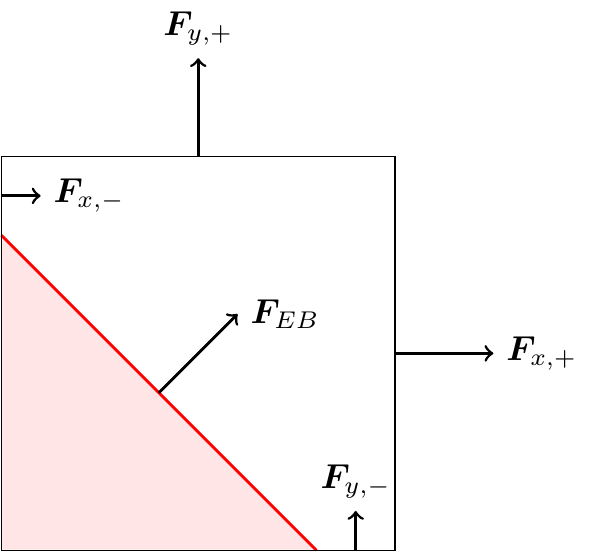}
    \caption
    {
        Decomposition of fluxes employed in the cell-averaged evaluation of divergence. Summation
        of these fluxes multiplied by cell surface areas leads to a conservative estimate of the
        divergence of the flux field. 
    }
    \label{fig:cut_cell_fluxes}
\end{figure}

Let us first consider terms which can be written as the divergence of a flux $\bm{F}$,
i.e.
\begin{equation}
    \bm{D} = \nabla \cdot \bm{F}.
\end{equation}
For these, we utilise the flux redistribution technique for embedded boundaries as
developed by Colella et al.~\cite{colella2006cartesian, graves2013cartesian}. Applying the
divergence theorem to a cut cell control volume, we find that
\begin{equation}
    \bm{D}^C_\bi
    = \frac{1}{V_\bi} \int_{\Omega_\bi} \nabla \cdot \bm{F} \diff V
    = \frac{1}{V_\bi} \int_{\partial \Omega_\bi} \bm{F} \cdot \diff \bm{A}
    = \frac{1}{V_\bi} \sum_{f=1}^{N_\bi} A_f \bm{F} \cdot \bn_f
    \label{eq:divergence_conservative_general}
\end{equation}
is a conservative estimate of $\bm{D}$. Here, $V_\bi = \alpha_\bi \Delta x \Delta y \Delta z$ is
the cell volume, $N_\bi$ is the number of faces (EB or otherwise) enclosing the cell, and $A_f$ is
the uncovered surface area of the face $f$, while $\bn_f$ is its normal vector. For all non-EB
faces, we can evaluate the flux tensor $\bm{F}$ just as for uncovered cells, since physical
variables are stored at cell centres. Consequently, the slope computation and upwinding procedures
do not need to be altered for EB cells except that one-sided upwinding is applied in the case of
covered neighbouring cells. In other words, we just need to use the EB information to extrapolate
$\bm{F}$ to the uncovered face centroid and multiply it by the corresponding face area.
Subscripting fluxes and area fractions in direction $d$ by $+$ at one end and $-$ at the other,
\eqref{eq:divergence_conservative_general} can be written
\begin{equation}
    \bm{D}^C_\bi
    = \frac{1}{V_\bi}
    \sum_{d \in \{ x,y,z \}} \left( \bm{F}_{d,+} A_{d,+} - \bm{F}_{d,-} A_{d,-} \right) \cdot
    \hat{\bm{e}}_d
    + \frac{1}{V_\bi} \bm{F}_{EB} A_{EB} \cdot \bn_{EB},
    \label{eq:divergence_conservative}
\end{equation}
and the only special consideration required is the evaluation of the flux tensor at the EB
centroid, $\bm{F}_{EB}$. Figure \ref{fig:cut_cell_fluxes} displays these fluxes in a 2D slice of
the cut cell. 

The downside of the conservative flux as computed in \eqref{eq:divergence_conservative} is that the
so-called small cell problem arises in the explicit temporal discretisation
\cite{leveque1988high, leveque1988cartesian}. The time step size is restricted since it is
proportional to $\alpha$, which can be arbitrary small in cut cells. In order to circumvent this,
we define a non-conservative approximation to the divergence as an average of the conservative
approximations in the neighbourhood of the cell,
\begin{equation}
    \bm{D}^{NC}_\bi
    = \frac{\sum\limits_{\bi' \in \mathcal{N}(\bi)} \alpha_{\bi'} \bm{D}^C_{\bi'}}
    {\sum\limits_{\bi' \in \mathcal{N}(\bi)} \alpha_{\bi'}} .
    \label{eq:divergence_nonconservative}
\end{equation}
We define the neighbourhood as the set
\begin{equation}
    \mathcal{N}(\bi) =
    \left\{
        \bi' \in \mathbb{Z}^3 : \min \lvert \bi - \bi' \rvert = \max \lvert \bi - \bi' \rvert = 1 ,
        \alpha_{\bi'} > 0
    \right\} ,
    \label{eq:neighbourhood}
\end{equation}
i.e.~all uncovered or cut cells whose index vector components differ by at most one from $\bi$,
except cell $\bi$ itself. A linear hybridisation of the conservative and non-conservative flux
approximations is then given by
\begin{equation}
    \bm{D}^H_\bi
    = \alpha_\bi \bm{D}^C_\bi + (1 - \alpha_\bi) \bm{D}^{NC}_\bi .
    \label{eq:divergence_hybrid}
\end{equation}
Note that this hybrid flux approximation has the desired behaviour in the limits of $\alpha_\bi \in
[0,1]$, and removes the effect of the local volume fraction on the time step size.

Although the hybrid flux stabilises the CFL restriction, it does not strictly enforce conservation.
The excess material lost or gained due to the usage of \eqref{eq:divergence_hybrid} rather than
\eqref{eq:divergence_conservative} is
\begin{equation}
    \delta \bm{D}_\bi
    = \alpha_\bi \left( \bm{D}^C_\bi - \bm{D}^H_\bi \right)
    = \alpha_\bi (1 - \alpha_\bi) \left( \bm{D}^C_\bi - \bm{D}^{NC}_\bi \right) ,
    \label{eq:divergence_excess}
\end{equation}
and this excess must be redistributed back to the cell neighbours. This is done with the use of
weights ${w_{\bi,\bi'} \geq 0}$ which quantify the fraction of $\delta \bm{D}_\bi$ redistributed to
cell $\bi'$, and which must satisfy ${\sum_{\bi' \in \mathcal{N}(\bi)} w_{\bi,\bi'} \alpha_{\bi'} =
1}$ for strict conversation. We use the simple weights
\begin{equation}
    w_{\bi,\bi'} = \frac{1}{\sum\limits_{\bi' \in \mathcal{N}(\bi)} \alpha_{\bi'}} ,
    \label{eq:weights}
\end{equation}
which are actually independent of $\bi'$. The final, discrete approximation to the divergence
operator is thus
\begin{equation}
    \bm{D}_\bi = \bm{D}^H_\bi + \sum_{\bi' \in \mathcal{N}(\bi)} w_{\bi',\bi} \delta \bm{D}_{\bi'}
    .
    \label{eq:divergence_final}
\end{equation}

\subsubsection{Convective term}

By the chain rule, we have
\begin{equation}
    \nabla \cdot \left( \bu \otimes \bu^F \right) = \bu \cdot \nabla \bu^F + \left( \nabla \cdot
    \bu \right) \bu^F = \bu \cdot \nabla \bu^F ,
\end{equation}
where the last equality holds due to the incompressibility constraint in
\eqref{eq:governing:incompressibility}. Seeing the convective term as the divergence of a tensor
flux, $\bm{C} = \nabla \cdot \bm{F}^{\bm{C}}$, we can apply the procedure outlined above in order to
compute the convective term in EB cells. Note that we enforce an inhomogeneous Dirichlet condition
on $\bm{u}$ at all EB walls, so that the components of the convective flux tensor are all zero
there: $\bm{F}^{\bm{C}}_{EB} = \bu \otimes \bu^{EB} = 0$. As such, we do not need to make any special
considerations for the convective fluxes at embedded boundaries. This is in contrast to the viscous
wall fluxes, which we will deal with next.

\subsubsection{Viscous term}

\begin{figure}
    \centering
    \includegraphics[width=\columnwidth]{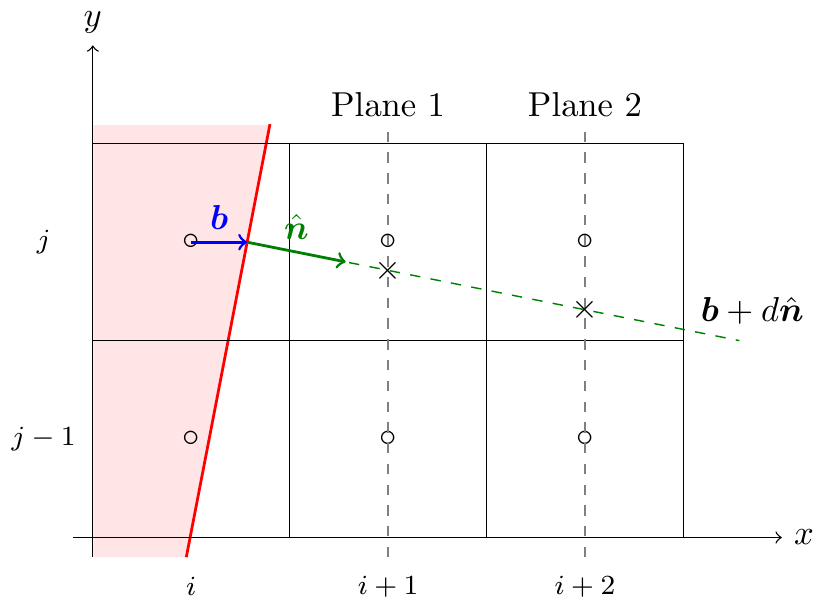}
    \caption
    {
        Computation of viscous wall fluxes at the EB centroid of cell $i,j,k$. The red line
        represents the embedded boundary, which cuts the cell such that its surface centroid and
        normal are $\bm{b}$ and $\bn$, respectively. In this example, the largest component of the
        surface normal is in the $x$-direction, so we interpolate to the points marked with crosses
        in the $yz$-planes corresponding to $i+1$ and $i+2$.
    }
    \label{fig:viscous_wall_flux}
\end{figure}

The flux tensor arising from the viscous term is ${\bm{F}^{\bm{V}} = \eta \left( \nabla \bu
\right)^\top}$, which can be non-zero on the EB surfaces. We therefore need a procedure to compute
the viscous wall flux at the EB surface centroid $\bm{b}$, which is given relative to local cut
cell coordinates, where the cell centre is the origin. To this end, we compute the gradient of each
velocity component along the EB surface normal vector $\bn_{EB}$. In order to achieve this, we
utilise AMReX' built-in biquadratic interpolation routine to find the value of $\bu$ at two points
located at distances $d_1$ and $d_2$ from the EB surface centroid $\bm{b}$. Figure
\ref{fig:viscous_wall_flux} illustrates the interpolation points for an example cut cell.

We start by finding which is the largest component of the surface normal vector. The biquadratic
interpolation will be done in planes where the corresponding coordinate is held fixed. Consider the
case when $\max ( n_x, n_y, n_z ) = n_x$, as in figure \ref{fig:viscous_wall_flux}. In order to
make sure that we are moving away from the EB, we let $s = \mathrm{sign} (n_x)$ and define the
interpolation points as those where the line $\bm{b} + d \bn$ intersect the planes $x = s$ and
$x=2s$. The corresponding distances from $\bm{b}$ are
\begin{equation}
    d_1 = \frac{s - b_x}{n_x} , \quad d_2 = \frac{2s - b_x}{n_x} .
    \label{eq:interpolation_distances}
\end{equation}

We can thus find the $y$ and $z$ coordinates, and utilise biquadratic interpolation to obtain
velocity values based on the 9 nearest points in the plane. Denoting the interpolated velocities by
$\bu_1$ and $\bu_2$, and allowing a prescribed velocity $\bu_{EB}$ on the boundary (zero throughout
this paper), \eqref{eq:lagrange_polynomial_gradient} gives the normal derivative as
\begin{equation}
    \left. \frac{\partial \bu}{\partial \bn} \right\vert_{EB}
        = \frac{d_2^2 (\bu_1 - \bu_{EB}) - d_1^2 (\bu_2 - \bu_{EB})}{d_1 d_2 (d_2 - d_1)}
        = \frac{d_2^2 \bu_1 - d_1^2 \bu_2}{d_1 d_2 (d_2 - d_1)} .
    \label{eq:dudn}
\end{equation}

All components of the velocity gradient are now available by taking the projections of $\partial
\bu / \partial \bn$ in the relevant Cartesian direction, i.e.~multiplying by the corresponding
component of $\bn$. We can thus compute $\strainrate$, $\eta$ and finally $\bm{F}^V = \eta \left(
\nabla \bu \right)^\top$ in the cut cell.

\subsubsection{Strain rate tensor and apparent viscosity}

Since our aim is to accurately capture the flow patterns of fluids with apparent viscosity
functions which depend strongly on the magnitude of the rate-of-strain tensor, it is essential that
we can compute the components of the velocity gradient tensor to second-order accuracy. The
procedure outlined for the viscous wall fluxes above is tailored for computing the values on EB
walls, but in the present case we require them at cell centres. In order to circumvent the problem
of covered neighbour cells, we adjust the stencil used for difference estimation. Similarly to in
the previous subsection, we consider a function $y(x)$ whose values we know at three points $x_i, i
\in \{0,1,2\}$, with the alteration that $x_0 \neq 0$ but the points have equal spacing $\Delta x$.
According to \eqref{eq:lagrange_polynomial_gradient}, the gradient of a quadratic polynomial fit to
these points is given by
\begin{align}
    \nonumber
    \frac{\partial f}{\partial x} = \frac{1}{\Delta x^2}
    & \left( \half (x-x_1 + x-x_2) y_0 \right. \\
    \nonumber
    &- (x-x_2 + x-x_0) y_1 \\
    &+ \left. \half (x-x_0 + x-x_1) y_2 \right) .
\end{align}
We can thus evaluate the gradient of the function at the given points to second order accuracy with
simple coefficients, since
\begin{subequations}
\begin{align}
    \left. \frac{\partial f}{\partial x} \right\vert_{x_0}
        &= \frac{1}{\Delta x} \left( - \frac{3}{2} y_0 + 2 y_1 - \half y_2 \right) \\
    \left. \frac{\partial f}{\partial x} \right\vert_{x_1}
        &= \frac{1}{2 \Delta x} \left( y_2 - y_0 \right), \\
    \left. \frac{\partial f}{\partial x} \right\vert_{x_2}
        &= \frac{1}{\Delta x} \left( \half y_0 - 2 y_1 + \frac{3}{2} y_2 \right) .
\end{align}
\end{subequations}

In order to evaluate the velocity gradient tensor in cut cells, we simply check whether the
neighbouring cells in each direction are covered, and if so, utilise a one-sided quadratic
difference estimator rather than the central one. Consequently, we are only fishing for values in
well-defined cells. With the estimates for velocity gradient components in place, we can evaluate
the cell-centred rate-of-strain magnitude and apparent viscosity directly.


\section{Verification}
\label{sec:verification}

Before evaluating the methodology for genuinely three-dimensional viscoplastic flows, we need to
ensure that the underlying incompressible flow solver has the desired order of accuracy by
performing grid convergence studies for problems with known solution. Furthermore, we verify that
the embedded boundaries work as expected by computing the solution to a Bingham Poiseuille flow in
a cylinder. 

\subsection{Spatio-temporal convergence study: Taylor-Green vortices}

In order to demonstrate spatio-temporal second order convergence of the presented algorithm, we
consider the unsteady Taylor-Green vortices in a Newtonian fluid, for which an
analytical solution exists. The computational domain is $\Omega = [0, \mathcal{L}]^2 \times [0,
\mathcal{W}]$, periodic boundary conditions are applied on all faces, and we take $\mathcal{U}$ as
the maximum of the initial velocity distribution. We obtain dimensionless variables by scaling
lengths by $\mathcal{L}$, velocities components by $\mathcal{U}$, time by $\mathcal{L} /
\mathcal{U}$ and pressure by the kinetic energy density $\rho \mathcal{U}^2$. The Taylor-Green
solution is then
\begin{subequations}
\label{eq:taylor-green}
\begin{align}
    \hat{u} &=
    \sin \left( 2 \pi \hat{x} \right)
    \cos \left( 2 \pi \hat{y} \right)
    e^{-8 \pi^2 \hat{t} / \Rey} , \\
    \hat{v} &=
    -\cos \left( 2 \pi \hat{x} \right)
    \sin \left( 2 \pi \hat{y} \right)
    e^{-8 \pi^2 \hat{t} / \Rey} , \\
    \hat{p} &= \frac{1}{4} \left( \cos \left( 4 \pi \hat{x} \right) + \cos \left( 4 \pi \hat{y} \right) \right)
    e^{-16 \pi^2 \hat{t} / \Rey} .
\end{align}
\end{subequations}
where we have introduced the Reynolds number $\Rey = \rho \mathcal{U} \mathcal{L} / \mu$.

For a series of spatial resolutions, characterised by the amount of cells, $N$, in each direction,
our system is advanced to the time ${\hat{t} = 1}$ with $\Rey = 100$. The resulting velocity field
is subtracted from the analytical solution in each cell, and we denote this residual
$\varepsilon_N(\bx)$. For two meshes with resolution $N$ and $2N$, the convergence rate of our
numerical method can be computed as
\begin{equation}
    r_* = \log_2 \norm{\frac{\varepsilon_{N}}{\varepsilon_{2N}}}_*
    \label{eq:convergence-rate}
\end{equation}
where $\norm{\cdot}_*$ is an appropriate function norm, typically one of
\begin{subequations}
\begin{align}
    &\norm{\varepsilon}_{1} = \int_{\Omega} |\varepsilon(\bx)| {\rm d} \bx , \\
    &\norm{\varepsilon}_{2} = \left( \int_{\Omega} |\varepsilon(\bx)|^2 {\rm d} \bx
    \right)^\half , \\
    &\norm{\varepsilon}_{\infty} = \max_{\Omega} |\varepsilon(\bx)| .
    \label{eq:norms}
\end{align}
\end{subequations}

Table \ref{tab:convergence} shows how the convergence rate for our solver approaches two in the
discrete version of all these norms as $N$ grows, as expected. 

\begin{table*}[htbp]
    \centering
    \begin{tabular}{lSSSSSS}
        \toprule
        $N$ & {$\norm{\varepsilon}_{1}$} & {$r_1$} & {$\norm{\varepsilon}_{2}$} & {$r_2$} &
        {$\norm{\varepsilon}_{\infty}$} & {$r_{\infty}$} \\
        \colrule
        32 & 0.000446166 & {$-$} & 0.000568385 & {$-$} & 0.00127935 & {$-$} \\
        64 & 0.000184199 & 1.2763153482666674 & 0.000229894 & 1.3058996678775268 & 0.000472784 & 1.4361578879356245 \\
        128 & 5.22399e-05 & 1.8180411888603984 & 6.47553e-05 & 1.8278986312686611 & 0.000130564 & 1.8564240505614413 \\
        256 & 1.35806e-05 & 1.9436049149908419 & 1.67843e-05 & 1.9478859075224415 & 3.36766e-05 & 1.9549387660311537 \\
        512 & 3.45001e-06 & 1.9768747715013355 & 4.25775e-06 & 1.9789492223212408 & 8.55327e-06 & 1.9771985036152346 \\
        \botrule
    \end{tabular}
    \caption
    {
        Error norms and convergence rates for viscous incompressible flow in the Taylor-Green
        vortex.
    }
    \label{tab:convergence}
\end{table*}

\subsection{Convergence study for viscoplastic fluid}
In order to evaluate the convergence of our code for viscoplastic fluids, we consider plane
Poiseuille flow between parallel plates separated by a width $2\mathcal{W}$, with the plane $z=0$
lying halfway between them. Poiseuille flow is driven by a constant pressure gradient $\mathcal{G}$
applied in the direction of flow, which is aligned with the parallel plates. Scaling distances by
$\mathcal{W}$ and velocity by the maximum velocity $\mathcal{U}$ of an unregularised Bingham fluid,
the analytical solution at steady-state is 
\begin{equation}
    \hat{u} (\hat{z}) = \begin{cases}
        1 , &\quad 0 \leq |\hat{z}| \leq z_0 , \\
        1 - \left( \frac{|\hat{z}| - z_0}{1 - z_0} \right)^2, &\quad z_0 < |\hat{z}| \leq 1 .
    \end{cases} 
    \label{eq:poiseuille:bingham}
\end{equation}
Here, we have introduced the dimensionless length 
\begin{equation}
    z_0 = \frac{\tau_0}{\mathcal{G} \mathcal{W}} ,
    \label{eq:poiseuille_z0}
\end{equation}
which separates the flow into yielded and unyielded regions, i.e.~represents the yield surface.
Note that since we have chosen to fix the maximum dimensionless velocity to unity, $z_0$ is the only
free variable in the system, and represents the relative strength of the yield stress
to the applied pressure gradient. For the sake of completeness, we note that the characteristic
velocity is given by
\begin{equation}
    \mathcal{U} = \frac{\tau_0 \mathcal{W} \left( 1 - z_0 \right)^2}{2 \mu z_0}  .
    \label{eq:poiseuille_umax}
\end{equation}
Although \eqref{eq:poiseuille_z0} gives us the analytical solution for a Bingham fluid, it would
not be correct to use for convergence analysis since the Papanastasiou regularisation which we are
employing is not an equivalent formulation. We have therefore derived the analytical solution for a
Papanastasiou regularised Bingham fluid, as shown in appendix \ref{app:papabing}. It is given by 
\begin{widetext}
\begin{equation}
    \hat{u} (\hat{z}) = 1 - \left( \frac{\hat{z} - z_0}{1 - z_0} \right)^2    
    + \frac{z_0 \varepsilon}{2 \xi}
    \left( 
    \left( 1 + W \left( \xi e^{-\xi (\hat{z}/z_0 - 1)} \right) \right)^2 
    - \left( 1 + W \left( \xi e^{-\xi (1/z_0 - 1)} \right) \right)^2 
    \right) ,
    \label{eq:poiseuille:papabing}
\end{equation}
\end{widetext}
where $\xi = \frac{2 z_0}{\varepsilon (1 - z_0)^2}$, and where $W(x)$ is the Lambert $W$ function,
which is also known as the product logarithm, since it is defined as the solution to $x = W(x
e^x)$. 

\begin{table*}[htbp]
    \centering
    \begin{tabular}{llSSSSSS} 
        \toprule
        $z_0$ & $N$ & {$\norm{\varepsilon}_{1}$} & {$r_1$} & {$\norm{\varepsilon}_{2}$} & {$r_2$} &
        {$\norm{\varepsilon}_{\infty}$} & {$r_{\infty}$} \\ 
        \colrule
        0   & 
        16  & 0.00390625 & {$-$} 
            & 0.00390625 & {$-$} 
            & 0.00390625 & {$-$} \\ &
        32  & 0.0009765625 & 2.0 
            & 0.0009765625 & 2.0
            & 0.0009765625 & 2.0 \\ &
        64  & 0.000244140625 & 2.0
            & 0.000244140625 & 2.0
            & 0.000244140625 & 2.0 \\ &
        128 & 6.103515625e-05 & 2.0
            & 6.103515625e-05 & 2.0
            & 6.103515625e-05 & 2.0 \\
        \colrule 
        0.1 & 
        16  & 0.004047297123504257   & {$-$} 
            & 0.004994331492935633   & {$-$} 
            & 0.010515282169174966   & {$-$} \\ &
        32  & 0.001233156197487274   & 1.714603215211083
            & 0.0016862240958587578  & 1.566495301298259
            & 0.003846973944267029   & 1.4506916008152209 \\ &
        64  & 0.00035055621511309674 & 1.814637831857419
            & 0.0005260720255991414  & 1.6804640398611264
            & 0.0013134550306872494  & 1.5503572510674763 \\ &
        128 & 9.55736675266982e-05   & 1.8749607265471435
            & 0.0001522792972406016  & 1.7885405182939529
            & 0.00041323122452441297 & 1.6683456309889535 \\ &
        256 & 2.513653011839418e-05  & 1.9268276694976763
            & 4.141130366966495e-05  & 1.8786232910487701
            & 0.0001182361570289947  & 1.8052779868987598 \\
        \colrule
        0.2 & 
        16  & 0.010176678398943928   & {$-$} 
            & 0.014165628688412963   & {$-$} 
            & 0.024892179293967143   & {$-$} \\ &
        32  & 0.003331050911760799   & 1.6112174426135635
            & 0.005012265942812364   & 1.4988597628627593
            & 0.009446229577178222   & 1.3978820391739075 \\ &
        64  & 0.0010134282311891837  & 1.7167334804645695
            & 0.0016088150511440788  & 1.6394644799411604
            & 0.0031819747962900102  & 1.5698161901163665 \\ &
        128 & 0.00029242028731112394 & 1.7931286137892901
            & 0.0004812822318994311  & 1.7410434162263366
            & 0.00102698535119905    & 1.631506805083072 \\ &
        256 & 8.063588951207906e-05  & 1.8585494018250763
            & 0.00013572596342156738 & 1.826186437937227
            & 0.00031235553317798814 & 1.7171546117066931 \\ 
        \colrule
        0.5 & 
        16  & 0.1220255804161702    & {$-$} 
            & 0.1339828770440298    & {$-$} 
            & 0.1629063441171097    & {$-$} \\ &
        32  & 0.04199366096489659   & 1.538940142481592
            & 0.04803472868804964   & 1.4798988925726047
            & 0.0616699457169273    & 1.4014033067794092 \\ &
        64  & 0.01378160356601615   & 1.6074278029712599
            & 0.016339341984702122  & 1.5557279542764588
            & 0.022375125430922016  & 1.4626718065405153 \\ &
        128 & 0.004375471553379288  & 1.655233350793157
            & 0.0053356064346785454 & 1.6146257252739789
            & 0.0076402046405712465 & 1.5502125838463983 \\ &
        256 & 0.001343243042472653  & 1.7037181420626915
            & 0.001672859285784711  & 1.6733361575310617
            & 0.002548515694780429  & 1.5839540429644419 \\ 
        \botrule
    \end{tabular}
    \caption
    {
        Error norms and convergence rates for plane Poiseuille flow of a Papanastasiou regularised
        Bingham fluid with $\varepsilon = 0.01$.
    }
    \label{tab:convergence:bingham}
\end{table*}

We emphasise that the newly derived analytical solution given by \eqref{eq:poiseuille:papabing}
takes into account the effect of the regularisation parameter on the analytical solution, so that
that factor is removed from the numerical algorithm in our convergence study. We therefore use a
moderate $\varepsilon = 0.01$, which in any case is small enough to capture the yield stress
effects on this simple problem, and vary the amounts of grid points $N = 2 \mathcal{W} / \Delta x$
over the channel width. The experiment is repeated for several values of $z_0$, since we expect the
condition number of the numerical problem to increase with the Bingham number. As seen in table
\ref{tab:convergence:bingham}, the numerical solution converges to \eqref{eq:poiseuille:papabing}
with increasing spatial accuracy for all values of $z_0$. Since the analytical solution is
parabolic for the Newtonian case, we achieve a perfect second order convergence rate, but is drops
to slightly lower than two as the yield stress increases. This is as expected, and the solution in
the unyielded region (which is undefined for Bingham fluids) converges slowest. Following the
convergence analysis of Olshanskii \cite{olshanskii2009analysis}, table
\ref{tab:convergence:yielded} shows the corresponding convergence rates when we only consider the
yielded region ($\hat{z} > z_0$), which are much better even for $z_0 = 0.5$. The $\infty$-norm is
the same, since it measures the maximum error throughout the domain, which is immediately outside
the yield surface in all cases. Although the strongly non-linear viscosity function poses a
significant numerical challenge, our algorithm performs well. Even more rapid convergence could
possibly be achieved through the use of convergence acceleration methods such as those discussed by
Housiadas \cite{housiadas2017improved}.

\begin{table*}[htbp]
    \centering
    \begin{tabular}{llSSSSSS} 
        \toprule
        $z_0$ & $N$ & {$\norm{\varepsilon}_{1}$} & {$r_1$} & {$\norm{\varepsilon}_{2}$} & {$r_2$} &
        {$\norm{\varepsilon}_{\infty}$} & {$r_{\infty}$} \\ 
        \colrule 
        0.1 & 
        16  & 0.003123299259837013   & {$-$} 
            & 0.003565221239243292   & {$-$} 
            & 0.006496731146296453   & {$-$} \\ &
        32  & 0.0008651533119889322  & 1.8520430907765617
            & 0.0010846698076453943  & 1.7167356776200555
            & 0.00301665160596587    & 1.10676591468124 \\ &
        64  & 0.0002592243967145151  & 1.7387543095407298
            & 0.0003847013164746564  & 1.4954452562824923
            & 0.0013134550306872494  & 1.1995812816456102 \\ &
        128 & 6.750769405135529e-05  & 1.941077657937412
            & 0.00010799558975247584 & 1.8327663688139002
            & 0.00041323122452441297 & 1.6683456309889535 \\ &
        256 & 1.6498994716411562e-05 & 2.032673815996815
            & 2.7255780592766483e-05 & 1.9863382537253371
            & 0.0001182361570289947  & 1.8052779868987598 \\
        \colrule
        0.2 & 
        16  & 0.005333701769510905   & {$-$} 
            & 0.008005378503829234   & {$-$} 
            & 0.024892179293967143   & {$-$} \\ &
        32  & 0.002017456449005338   & 1.402599629517403
            & 0.0034829398422987893  & 1.2006640619335938
            & 0.009446229577178222   & 1.3978820391739075 \\ &
        64  & 0.0005568065969959065  & 1.8572893209399761
            & 0.0010580582645516824  & 1.718886479609492
            & 0.0031819747962900102  & 1.5698161901163665 \\ &
        128 & 0.00013319790780491015 & 2.063604882638235
            & 0.0002761484335724105  & 1.937903226166033
            & 0.00102698535119905    & 1.631506805083072 \\ &
        256 & 3.4448174079007475e-05 & 1.9510720013689218
            & 7.508894947673473e-05  & 1.8787714307289662
            & 0.00031235553317798814 & 1.7171546117066931 \\ 
        \colrule
        0.5 & 
        16  & 0.08189230412244795   & {$-$} 
            & 0.09801602162188305   & {$-$} 
            & 0.15799182653427124   & {$-$} \\ &
        32  & 0.023658676072963805  & 1.7913585366366906
            & 0.031225725178927913  & 1.6502825132674102
            & 0.0616699457169273    & 1.3572104431682432 \\ &
        64  & 0.006447386694631248  & 1.8755829238306898
            & 0.009383615029765825  & 1.7345193450468936
            & 0.022375125430922016  & 1.4626718065405153 \\ &
        128 & 0.0017166696868698202 & 1.909102044773576
            & 0.002730149298475482  & 1.7811639796558365
            & 0.0076402046405712465 & 1.5502125838463983 \\ &
        256 & 0.0004516479029166405 & 1.9263420545494685
            & 0.0007759818780158549 & 1.8148849814486843
            & 0.002548515694780429  & 1.5839540429644419 \\ 
        \botrule
    \end{tabular}
    \caption
    {
        Error norms and convergence rates for plane Poiseuille flow of a Papanastasiou regularised
        Bingham fluid with $\varepsilon = 0.01$, using only the solution in the yielded region. 
    }
    \label{tab:convergence:yielded}
\end{table*}


\section{Evaluation}
\label{sec:evaluation}

In order to properly evaluate the capabilities of our code, we need to simulate test cases which
exhibit fully three-dimensional effects which are characteristic of yield stress fluids. A widely
discussed case is that of bodies moving at constant speed through a Bingham fluid. Such bodies are
fully encapsulated by a so-called yield envelope, separating the unyielded bulk material from an
interior recirculating flow. This interior flow has a unique topology owing to the viscoplastic
nature of the fluid, and has been widely studied for two-dimensional \cite{ansley1967motion,
beris1985creeping, blackery1997creeping, liu2002convergence, tokpavi2008very, chaparian2017yield,
koblitz2018viscoplastic} and axisymmetric three-dimensional \cite{prashant2011direct,
chen2016three} shapes.

\subsection{Cylinder moving through Bingham fluid}

Although creeping flow past a sphere in a viscoplastic fluid was investigated first
\cite{ansley1967motion}, the simpler case is that of an infinitely long circular cylinder. This is
due to the fact that only a thin slice in 3D is required for comparisons with two-dimensional
reference results, but also because the viscoplastic effects are more dominant due to the extended
geometry. Consequently, there is no ambiguity regarding the resulting shape of yield surfaces. 
The problem was first considered by Adachi and Yoshioka
\cite{adachi1973creeping}, who used variational principles and slip-line analyses to estimate the
relevant drag effects. Randolph and Houlsby \cite{randolph1984limiting} used plasticity theory to
investigate the same problem in the context of pile pressure in cohesive soil. Two excellent
references which utilised numerical simulations to explore the problem are the papers by
Mitsoulis \cite{mitsoulis2004creeping} and Tokpavi et al.~\cite{tokpavi2008very}, in which the flow
around a highly resolved 2D quadrant surrounding the cylinder was simulated with Papanastasiou
regularisation schemes. The former investigated the influence of wall effects for creeping flow,
and found that it was small for medium to large Bingham numbers, while the latter extended the
parameter ranges and analysis of the problem. They showed that the occurrence of characteristic
dips in the yield envelope fore and aft of the cylinder is clearly prominent for Bingham fluids, in
addition to small unyielded caps attached to the poles of the cylinder and rigidly rotating
unyielded plugs in its equatorial plane. The same has been demonstrated by Chaparian and Frigaard
\cite{chaparian2017yield} by using an augmented Lagrangian approach, which captures the yield
surface without regularisation. They also show how slip line field theory captures the yield
envelope and polar caps well, but is ill-suited for the equatorial plugs. It is also worth
mentioning that there have been a number of studies with multiple cylinders in various collinear
configurations, all showing similar large-scale features \cite{liu2003interactions,
prashant2011direct, koblitz2018viscoplastic}. Since the problem is essentially two-dimensional due
to its planar symmetry, we expect to see the same yield surface shapes as the aforementioned
authors, even though our code is fully three-dimensional.

We thus consider an infinitely long circular cylinder, where the direction of flow (along the unit
vector $\hat{\bm{e}}_z$) is perpendicular to the cylinder axis ($\hat{\bm{e}}_y$). The
computational domain is $\Omega = [0, 4 \mathcal{D}] \times [0, \mathcal{L}] \times [0, 6
\mathcal{D}]$, where $\mathcal{L}$ is the length of the computational domain along the cylinder and
$\mathcal{D}$ is its diameter, which we take as the characteristic length scale. In the following
simulations, we have used $\mathcal{L} = \mathcal{D}/2$ and periodic boundary conditions in the
$y$-direction. Rather than imposing a constant velocity on the cylinder, we keep it fixed and
instead let $\bm{u} = \mathcal{U} \hat{\bm{e}}_z$ everywhere initially and at the domain boundaries
as the simulation progresses. This means that the cylinder is falling with the same speed
($\mathcal{U}$) in the reference frame where the bulk fluid is at rest. Periodic boundary
conditions are imposed in the axial direction of the cylinder.

Apart from the regularisation parameter $\varepsilon$, there are two dimensionless
numbers which govern the flow, namely the Reynolds number
\begin{equation}
    \Rey = \frac{\rho \mathcal{U} \mathcal{D}}{\mu} ,
\end{equation}
and the Bingham number
\begin{equation}
    \Bin = \frac{\tau_0 \mathcal{D}}{\mu \mathcal{U}} ,
\end{equation}
which quantifies the relative strength of yield stress to viscous stress. We are interested in the
creeping flow regime with a high degree of viscoplasticity, so we let $\Rey = 10^{-3}$.  As
numerical parameters, we let $N = \mathcal{D} / \Delta x = 64$ and $\varepsilon = 10^{-3}$.  We
expect these choices to be suitable for capturing the relevant flow features in this problem, based
on the work of Liu et al.~\cite{liu2002convergence} and our own results in subsection
\ref{subsec:sphere}.

For flow past bodies, quantitative comparisons with results from the literature are obtained
through the computation of a drag coefficient, which is the ratio of the total drag force $F_D$
on the body compared to a reference force. For the cylinder, we take the reference as the product
of the characteristic stress and the cross-sectional area of the cylinder $\mathcal{L}
\mathcal{D}$. Denoting the boundary of the body by $\mathcal{B}$, the drag force is given by the
surface integral 
\begin{equation}
    F_D = 
    \int_\mathcal{B} \hat{\bm{e}}_z \cdot \left( \stress - p \bm{I} \right) \cdot \diff \bm{S} ,
    \label{eq:drag-force}
\end{equation}
while the drag coefficient is 
\begin{equation}
    C_D = \frac {F_D} {\mu \mathcal{U} \mathcal{L}} .
    \label{eq:cylinder:drag-coef}
\end{equation}

\begin{figure}
    \centering
    \includegraphics[width=\columnwidth]{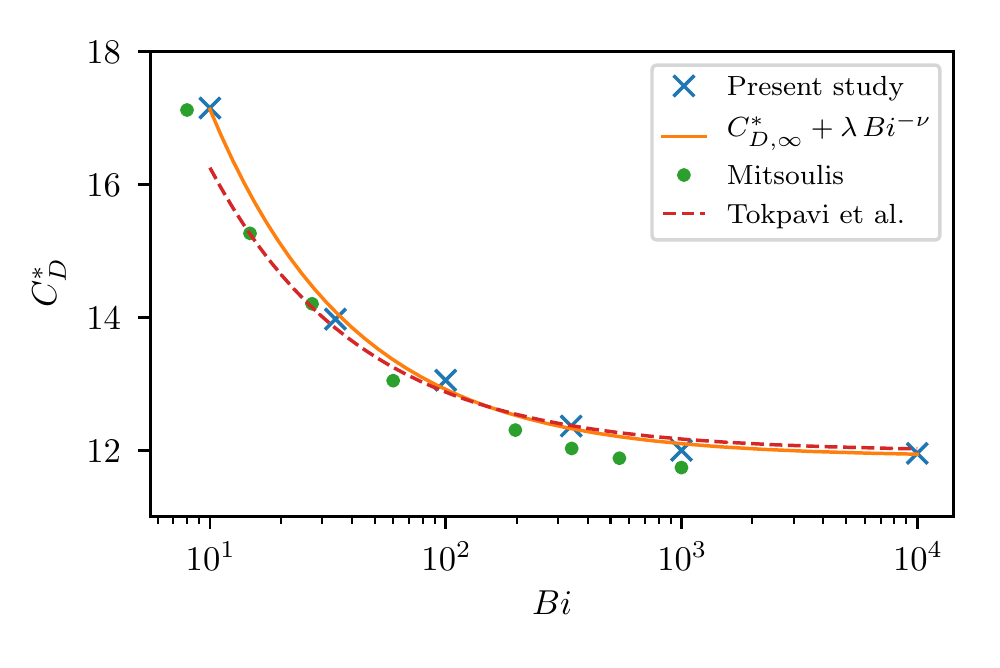}
    \caption
    {
        Quantitative comparison of our solution around the cylinder to that in the published
        literature \cite{mitsoulis2004creeping, tokpavi2008very} through computation of the drag
        coefficient at various Bingham numbers.
    }
    \label{fig:drag_cylinder}
\end{figure}

The integral in \eqref{eq:drag-force} is computed using numerical quadrature. In order to verify
the drag force calculations, we initially perform a simulation with a Newtonian fluid, resulting
in a computed drag coefficient equal to 21.8128. For comparison, Faxén's formula as given by
\eqref{eq:faxen} for a semi-analytical drag coefficient yields a value of 21.9642, i.e.~a relative
discrepancy of 0.69\%. By driving the system to steady-state for various Bingham numbers, we can
compare our results with others which have been published in the literature. Figure
\ref{fig:drag_cylinder} shows how the ratio $C_D^* = C_D / \Bin$ gradually approaches the
asymptotic value $C_{D, \infty}^*$ as $\Bin$ grows large. The numerical results of Mitsoulis
\cite{mitsoulis2004creeping} and Tokpavi et al.~\cite{tokpavi2008very} are overlain our own, and
the behaviour similar. Our results are closest to the trend line by Tokpavi et al., which they fit
to the function 
\begin{equation}
    C_D^* = C_{D,\infty}^* + \lambda \Bin^{-\nu} .
\end{equation}
Our data has been fit to the same function, and we have also computed the
gravity yield number 
\begin{equation}
    Y = \frac{\tau_0}{\frac{1}{4} \pi \mathcal{D} g \Delta \rho}
\end{equation}
where $g$ is the gravitational acceleration and $\Delta \rho$ is the density difference between the
solid cylinder and surrounding fluid. This dimensionless number is the smallest ratio of yield
stress to gravitational stresses which obstructs the solid from falling through the fluid, and for
the particular configuration at hand, we have $Y = 1 / C_{D, \infty}^*$
\cite{mitsoulis2004creeping}. Table \ref{tab:drag_cylinder} shows our quantitative results
alongside those by other authors. 
\begin{table}[htbp]
    \centering
    \begin{tabular}{lcccc}
        \toprule
        Reference                                       & $C_{D,\infty}^*$  & $Y$       & $\lambda$ & $\nu$     \\
        \colrule
        Adachi \& Yoshioka \cite{adachi1973creeping}    & 10.28             & 0.09728   &           &           \\
        Randolph \& Houlsby \cite{randolph1984limiting} & 11.94             & 0.08375   &           &           \\
        Mitsoulis \cite{mitsoulis2004creeping}          & 11.70             & 0.08547   &           &           \\
        Tokpavi et al.~\cite{tokpavi2008very}           & 11.98             & 0.08347   & 20.43     & 0.68      \\
        Present study                                   & 11.90             & 0.08403   & 26.99     & 0.7131    \\
        \botrule
    \end{tabular}
    \caption
    {
        Comparison of the yield number and limiting ratio of drag coefficient to Bingham number
        with various reference results from the literature. 
    }
    \label{tab:drag_cylinder}
\end{table}

\begin{figure*}
    \centering
    \begin{subfigure}{0.49\textwidth}
        \includegraphics[width=\columnwidth]{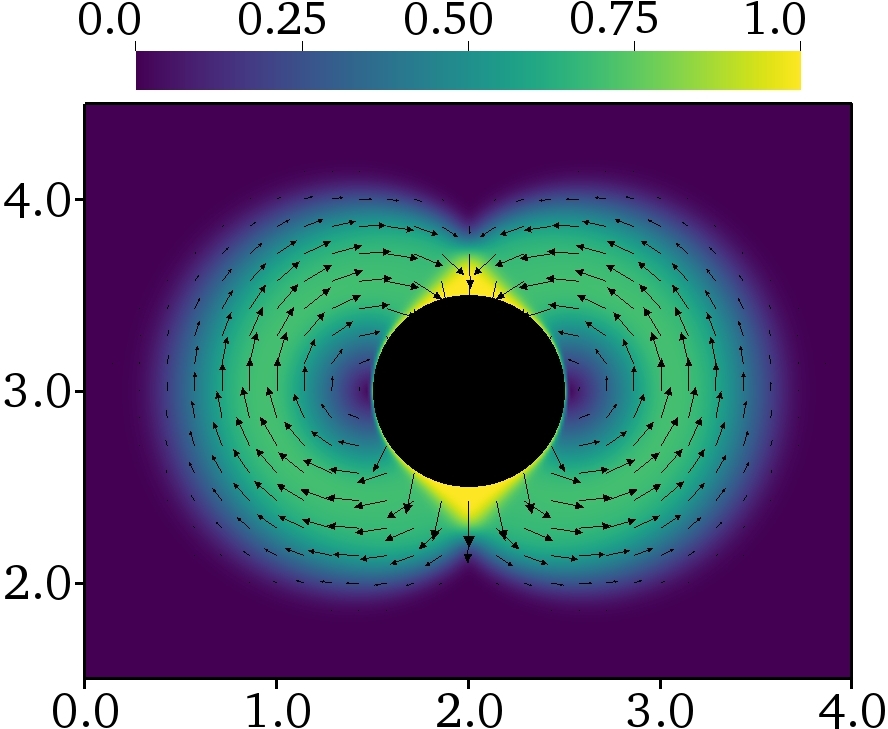}
        \caption
        {
            $\hat{\bm{e}}_z - \hat{\bu}$
        }
        \label{fig:cylinder_velocity}
    \end{subfigure}
    \begin{subfigure}{0.49\textwidth}
        \includegraphics[width=\columnwidth]{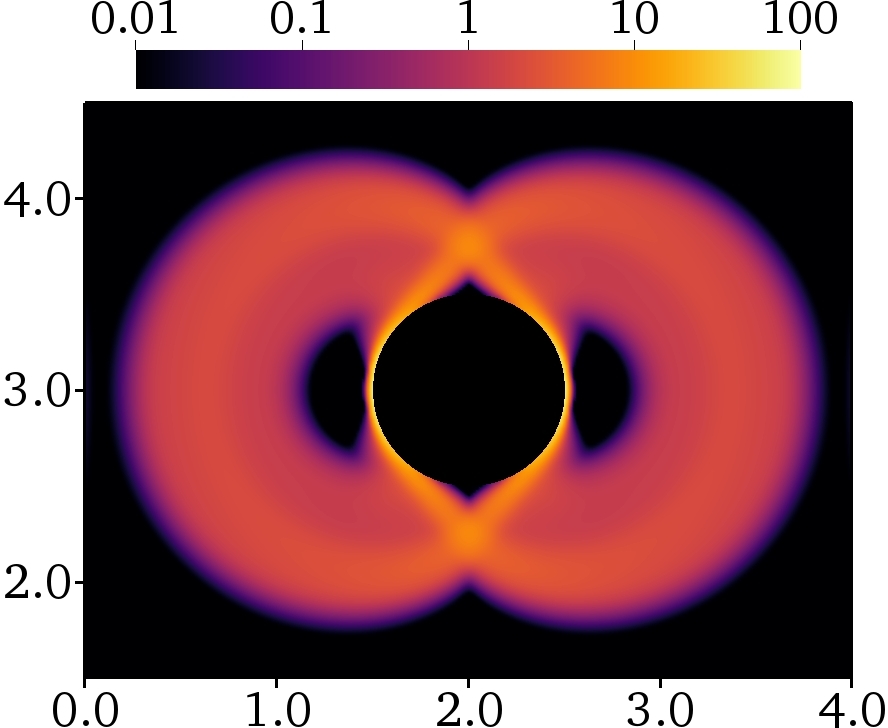}
        \caption
        {
            $\norm{\strainrate}$ (s$^{-1}$)
        }
        \label{fig:cylinder_strainrate}
    \end{subfigure}
    \\[2em]
    \begin{subfigure}{0.49\textwidth}
        \includegraphics[width=\columnwidth]{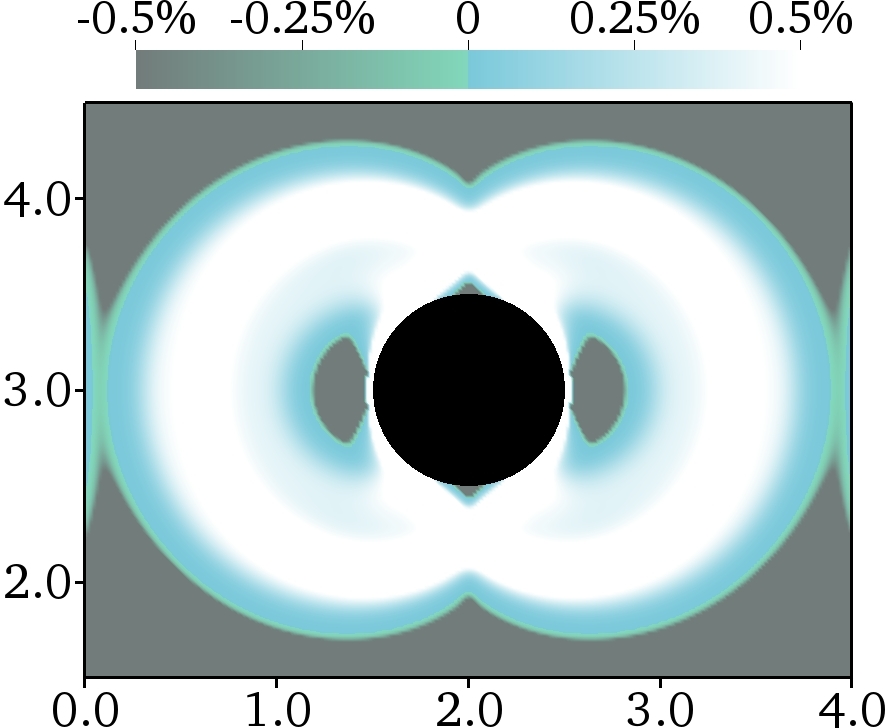}
        \caption
        {
            ${\norm{\stress} / \tau_0 - 1}$
        }
        \label{fig:cylinder_treskatis}
    \end{subfigure}
    \begin{subfigure}{0.49\textwidth}
        \includegraphics[width=\columnwidth]{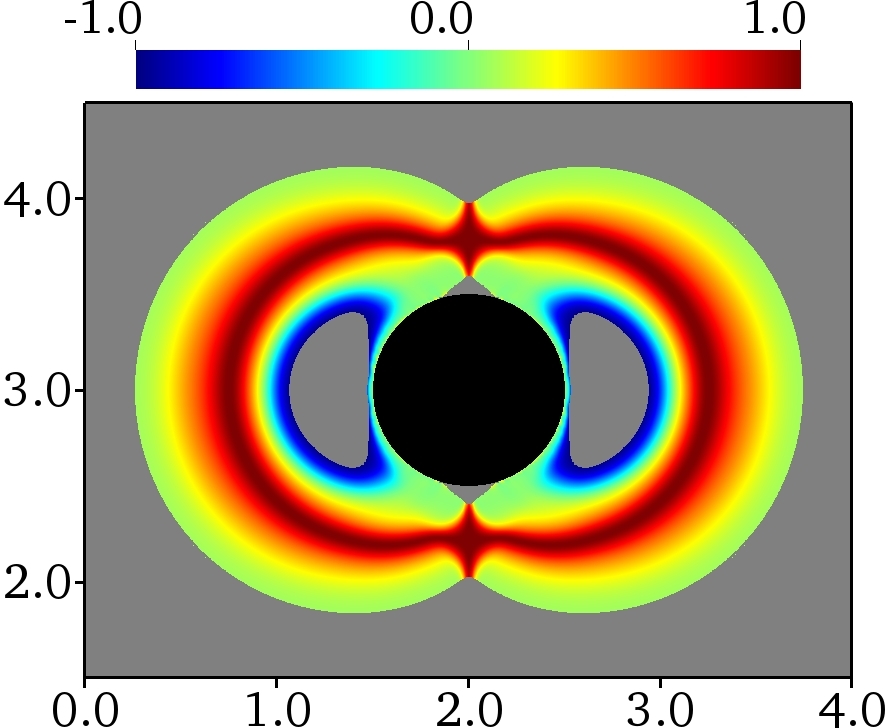}
        \caption
        {
            $\Lambda$
        }
        \label{fig:cylinder_ryovar}
    \end{subfigure}
    \caption
    {
        Flow characteristics in the $\hat{x}\hat{z}$-plane through the centre of the cylinder
        (black) at steady-state for $\Bin = 340.7$. Depicted are (a) the relative velocity field
        and its magnitude, (b) the rate-of-strain magnitude, (c) the yield surface stress deviation
        and (d) the flow topology parameter. Note that the colormap in (b) has logarithmic scaling,
        and that values outside the colormap for (c) are mapped on to the endpoints. In (d), the
        yield surface computed with $\delta = 10^{-3}$ is also masked out in grey. 
    }
    \label{fig:cylinder}
\end{figure*}

In addition to computing drag coefficients, there is a lot of insight to be gained from visualising
different metrics of the flow field. Figure \ref{fig:cylinder} therefore illuminates the resulting
flow field through plots of several different variables, for the case with $\Bin = 340.7$. The
Bingham number is chosen to match that in the study by Liu, Muller and Denn
\cite{liu2002convergence}, where it is stated that this number is large enough to diminish any
outer boundary wall effects for the given choice of domain size. The upper left plot, (a), shows
the relative speed with velocity vectors overlaid. A distinct boundary is visible, enclosing a
region of non-zero relative velocity surrounding the cylinder.  This boundary represents the yield
envelope of the body. The velocity vectors indicate a recirculating flow, sweeping material away
from the front of the travelling cylinder to the rear in a wide, circular arc. This results in
slowly moving material either side of the cylinder, in addition to material travelling at the same
speed as the cylinder at the polar caps. These observations are indicative of the expected
unyielded plugs rotating at the equator and clinging to the polar caps.

In the upper right plot, (b), the magnitude of the rate-of-strain tensor is shown with logarithmic
scaling for the colormap. This plot properly elucidates the low-strain regions of unyielded fluid,
and confirm the existence of rigidly rotating equatorial plugs and unyielded material attached to
the polar caps, as implied by the velocity distribution. Note that although the yield surface is
characterised by the contour $\norm{\strainrate} = 0$ for ideal Bingham plastics as given by
\eqref{eq:bingham:viscosity}, the regularised version \eqref{eq:papabing:viscosity} leads to a
yield surface given by
\begin{equation}
    \norm{\hat{\strainrate}} = \varepsilon W \left( \frac{\Bin}{\varepsilon} \right) ,
    \label{eq:lambertw}
\end{equation}
where we recall that $W(x)$ is the Lambert $W$ function.  With the current parameter values,
\eqref{eq:lambertw} gives a yield surface characterised by $\norm{\strainrate} = 0.010397$
s$^{-1}$, which is why we have used $10^{-2}$ s$^{-1}$ as the lower limit of the colormap.

In order to illustrate the location of the yield surface, an option would be to draw the contour
where the stress magnitude is equal to $\tau_0$.  However, as discussed previously in the
literature, there is instability near this stress value which means that a better measure of the
fully converged yield surface is the contour $\norm{\stress} = (1 + \delta) \tau_0$
\cite{olshanskii2009analysis}, where $\delta$ is some small parameter of the order $10^{-3}$. This
is because the solution converges much faster in the yielded region than in the unyielded ones. On
the other hand, Treskatis argues that a better visual investigation of the yield surface is
obtained by plotting the relative deviation from the yield surface, $\norm{\stress} / \tau_0 - 1$,
restricted to some small range around zero \cite{treskatis2016accelerated}. In this manner, we
avoid the introduction of systematic error through overestimation of the unyielded regions. Note
that this must be done using a colormap which changes abruptly at zero, as seen in figure
\ref{fig:cylinder_treskatis}. The transition is somewhat diffuse, but all the expected unyielded
regions (envelope, rotating plugs and polar caps) are clearly captured where the stress deviation
is negative. The actual yield surface is located somewhere in the transition from blue to white.

Finally, in figure \ref{fig:cylinder_ryovar} we have plotted a flow topology parameter which is a
normalised invariant of the velocity gradient tensor. It indicates the relative strengths of the
strain rate and the vorticity $\bm{\omega} = \nabla \times \bu \,$ \cite{davidson2015turbulence,
de2017viscoelastic},
\begin{equation}
    \Lambda =
    \frac{\norm{\strainrate}^2-\bm{\omega}^2}{\norm{\strainrate}^2+\bm{\omega}^2} .
    \label{eq:ryovar}
\end{equation}
This parameter accurately describes the nature of different parts of the flow regime, since values
of $\Lambda$ equal to $-1$, 0 and 1 correspond to pure rotation, shear and extension, respectively.
As expected, we observe pure rotation around the unyielded plugs on the equatorial line of the
cylinder. Adjacent to areas of rigid rotation, and near the cylinder, shear flow is evident. This
is also the case near the yield envelope. In fact, these two shearing regions represent two
distinct cases of viscoplastic boundary layers, as discussed originally by Oldroyd in 1947
\cite{oldroyd1947two} and more recently by Balmforth in his newly published lecture notes
\cite{balmforth2019viscoplastic}. We also point the interested reader to discussions surrounding
viscoplastic boundary layer around particles in some of the other works which have already been
mentioned \cite{beris1985creeping, tokpavi2008very, chaparian2017yield}. Finally, a belt of purely
extensional flow surrounds the cylinder, rotating plugs and polar caps. Note that in this plot, the
yield surfaces masked out in grey are computed with $\delta = 10^{-3}$. This allows us to verify
that we recover the same yield surface shapes as those computed in the references
\cite{tokpavi2008very, chaparian2017yield}, when using the same visualisation procedure. 

\subsection{Sphere moving through Bingham fluid}
\label{subsec:sphere}

Since the cylinder test case allows for direct comparisons with two-dimensional simulations, it is
a good test case for verifying the interplay between yield stress rheology and the embedded
boundaries. On the other hand, it does not exhibit any genuinely three-dimensional effects, and for
many real-world scenarios an infinite cylinder is not a realistic representation of the actual
bluff bodies. We therefore investigate the flow around a sphere in the same configuration,
retaining its diameter $\mathcal{D}$ as the characteristic length scale, but extending the domain
to $[0, 4\mathcal{D}]$ in the $y$-direction. This problem has been analysed by several authors
previously \cite{ansley1967motion, beris1985creeping, blackery1997creeping, liu2002convergence,
prashant2011direct, chen2016three}, but it still warrants further attention due to gaps relating to
important viscoplastic flow features. In particular, claims about how specific parts of the yield
surface depend on spatial resolution and regularisation parameter are not consistent. Furthermore,
few of them were based on three-dimensional simulations, and those that were tended to be hampered
by low spatial resolution, making it difficult to draw any conclusions regarding the shape and
extent of the yield surface.

Ansley and Smith first considered the problem in 1967, and proposed a yield surface based on slip
line field theory. This included dips in the yield envelope and polar caps \cite{ansley1967motion}.
Although it was a great contribution, they conceded that the qualitative shape of the yield surface
could not be supported by direct evidence. Beris et al.~phrased the problem as a free boundary
problem at the yield surface, and performed 2D simulations with a priori estimates of its location
\cite{beris1985creeping}. They captured the envelope dips and polar caps, and reported an
equatorial torus-shaped region with low strain rate. Its motion was described as close to the sum
of rigid translation in the direction of flow and solid body rotation, but they aptly noted that
perfectly rigid rotation is not possible in the configuration, except exactly in the equatorial
plane. This is due to the rotated strain field around the equator of the sphere. Liu et
al.~\cite{liu2002convergence} computed 2D results in a quadrant using Papanastasiou regularisation
and finite elements on a body fitted mesh. They reported the existence of both polar caps and a
equatorial torus, although the latter is unphysical according to the arguments put forth by Beris
et al. They showed that the plug regions shrank with decreasing regularisation parameter, but could
not infer the limiting behaviour. More recently, three-dimensional simulations in the framework of
lattice Boltzmann techniques have been published, utilising a dual viscosity model
\cite{prashant2011direct} and Papanastasiou regularisation \cite{chen2016three}. However, the
resolution for these simulations was low ($N = 12$), so it is difficult to draw accurate
conclusions regarding the yield surface shapes.

\begin{figure*}
    \centering
	\begin{subfigure}{0.3\textwidth}
		\includegraphics[width=\columnwidth]{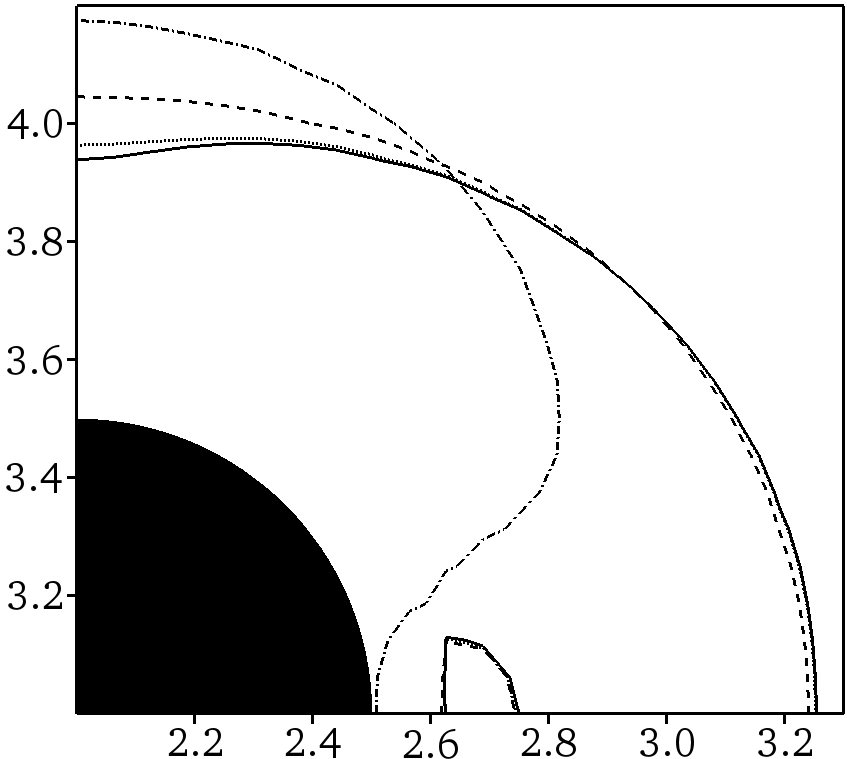}
        \caption{$N = 16$}
	\end{subfigure}
	~
	\begin{subfigure}{0.3\textwidth}
		\includegraphics[width=\columnwidth]{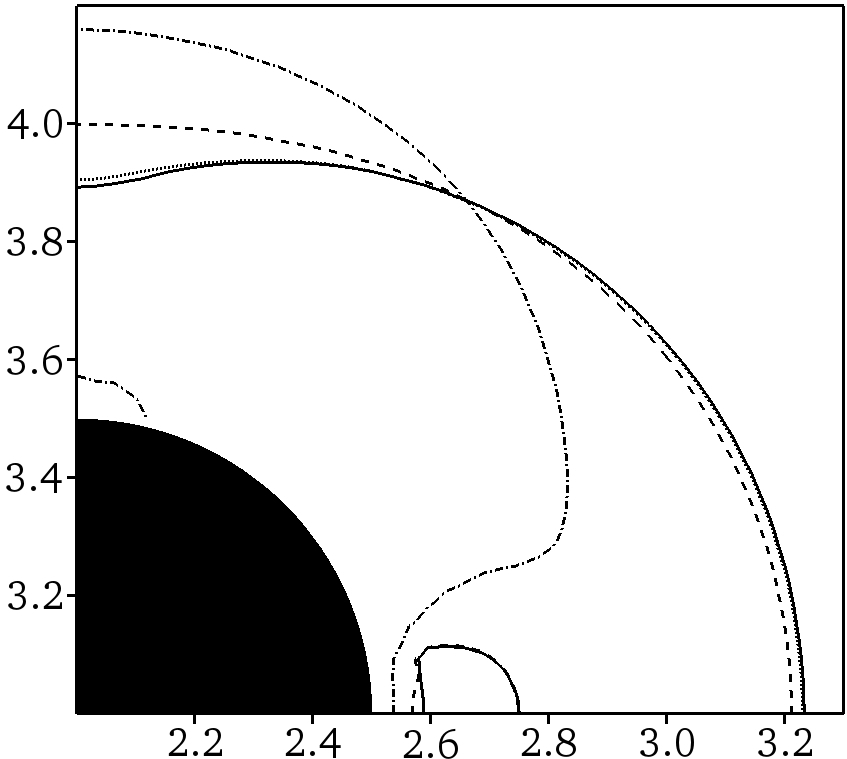}
        \caption{$N = 32$}
	\end{subfigure}
	~
	\begin{subfigure}{0.3\textwidth}
		\includegraphics[width=\columnwidth]{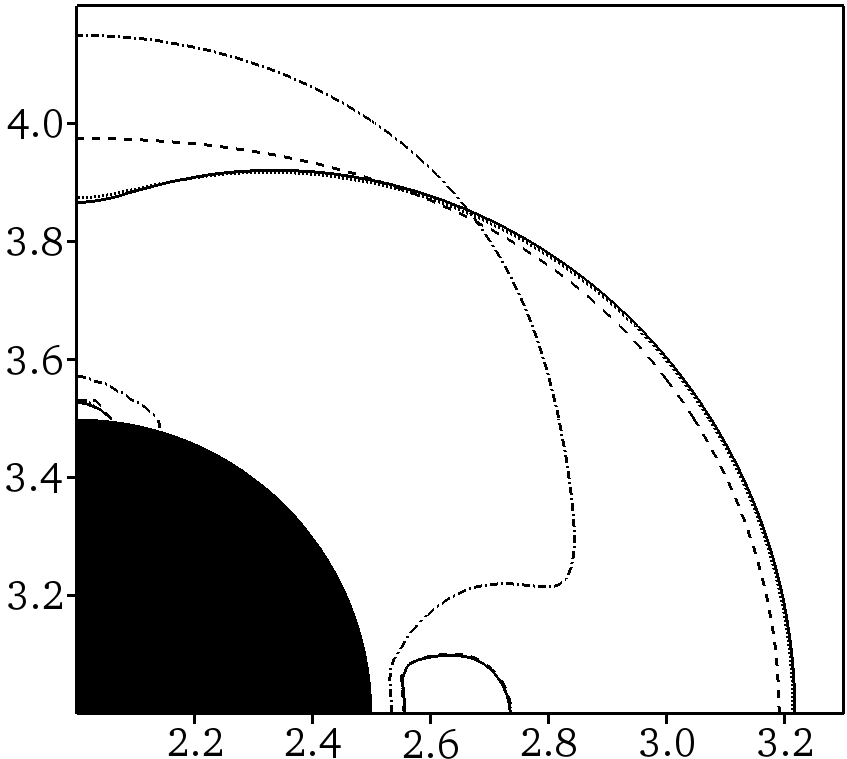}
        \caption{$N = 64$}
	\end{subfigure}
	\caption
	{
        Contour lines of the steady-state yield surface $\norm{\stress} = 1.001 \, \tau_0$ in the
        upper right corner of the $\hat{x}\hat{z}$-plane $\hat{y} = 2$, for a sphere translating in
        an infinite Bingham medium with $\Bin = 340.7$. The resolution of the numerical mesh
        employed in the simulations increases from (a) to (c). For each resolution, the different
        lines correspond to the following regularisation parameter values: \\
        \sampleline{thick, dash dot} $\varepsilon = 10^{-1}$ \hspace{1em} 
        \sampleline{thick, dashed} $\varepsilon = 10^{-2}$ \hspace{1em} 
        \sampleline{thick, dotted} $\varepsilon = 10^{-3}$ \hspace{1em}
        \sampleline{thick} $\varepsilon = 10^{-4}$
	}
	\label{fig:sphere_yieldsurface_slice}
\end{figure*}

The two parameters which determine the yield surface for a given Bingham number, are the
regularisation parameter $\varepsilon$ and the spatial resolution as given by $N$. Similarly to
what was done in the paper by Liu et al.~\cite{liu2002convergence}, in Figure
\ref{fig:sphere_yieldsurface_slice} we plot the computed yield surfaces in the upper right corner
of the plane $\hat{y} = 2$, which goes directly through the sphere, for a range of $\varepsilon$
values. However, we also consider three different mesh resolutions, in order to separate the
effects of the two parameters. It is clear that both parameters must have acceptable values in
order for the solution to converge as expected. Without a small enough regularisation parameter (at
least $\varepsilon = 10^{-3}$), the yield envelope dips and low-strain equatorial plugs are not
captured, whereas a low spatial resolution results in unresolved polar caps. We note that the
unphysical torus-shaped yield surface at the equator is apparent also in our simulations, but this
is due to the utilisation of $\delta = 10^{-3}$ in the visualisation of the yield surface, as
explained below. In the remainder of this section, we let $N=64$ and $\varepsilon = 10^{-3}$.

\begin{figure}
    \centering
    \includegraphics[width=\columnwidth]{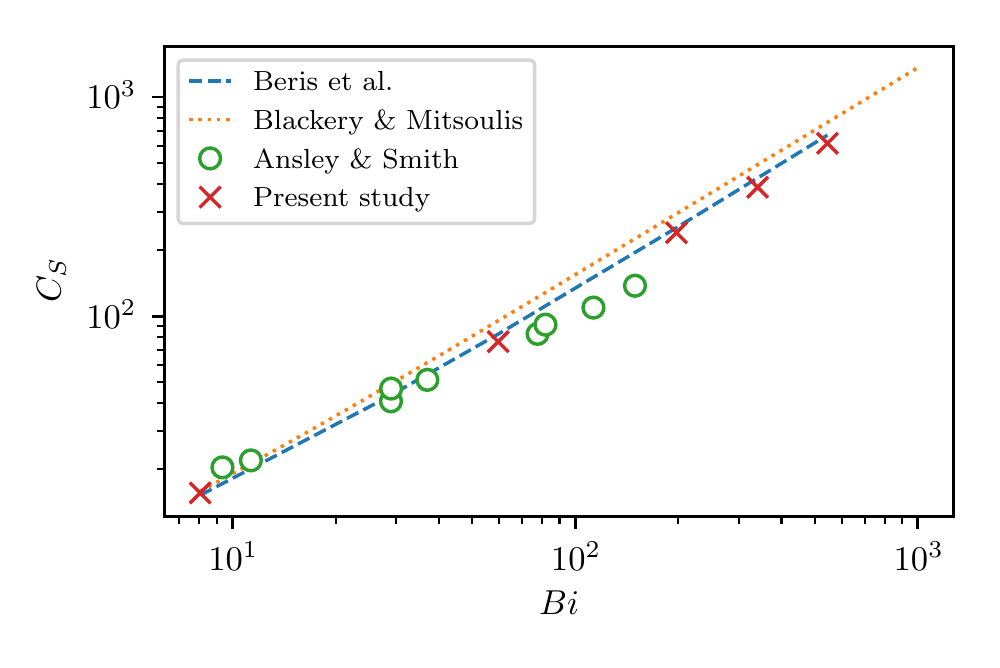}
    \caption
    {
        Quantitative comparison of our solution around the sphere to that in the published
        literature \cite{ansley1967motion, beris1985creeping, blackery1997creeping} through
        computation of the Stokes drag coefficient at various Bingham numbers. 
    }
    \label{fig:drag_sphere}
\end{figure}

For the case of a sphere, the reference force for the computation of the drag coefficient is the
Stokes force \cite{stokes1851effect} that acts on a sphere falling in an infinite Newtonian medium.
The Stokes drag coefficient is thus 
\begin{equation}
    C_S = \frac {F_D} {6 \pi \mu \mathcal{U} \mathcal{R}} .
    \label{eq:drag-coef}
\end{equation}
Figure \ref{fig:drag_sphere} shows our results alongside previously published data in the
literature. We have included the original computational results by Beris et
al.~\cite{beris1985creeping} and Blackery and Mitsoulis \cite{blackery1997creeping} as dashed and
dotted lines, respectively, and have also overlaid the experimental results of Ansley and Smith
\cite{ansley1967motion}. As is evident, our computed values for the Stokes drag coefficient lie
within the same range as the references, and match the results of Beris et al.~most closely.  

\begin{figure*}
    \centering
    \begin{subfigure}{0.49\textwidth}
        \includegraphics[width=\columnwidth]{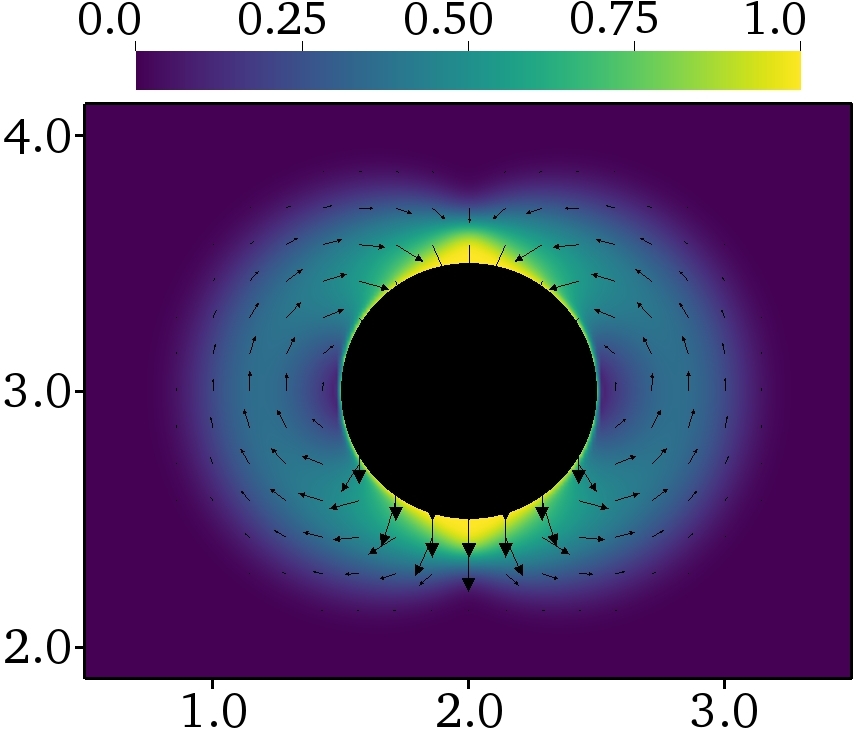}
        \caption
        {
            $\hat{\bm{e}}_z - \hat{\bu}$
        }
    \end{subfigure}
    \begin{subfigure}{0.49\textwidth}
        \includegraphics[width=\columnwidth]{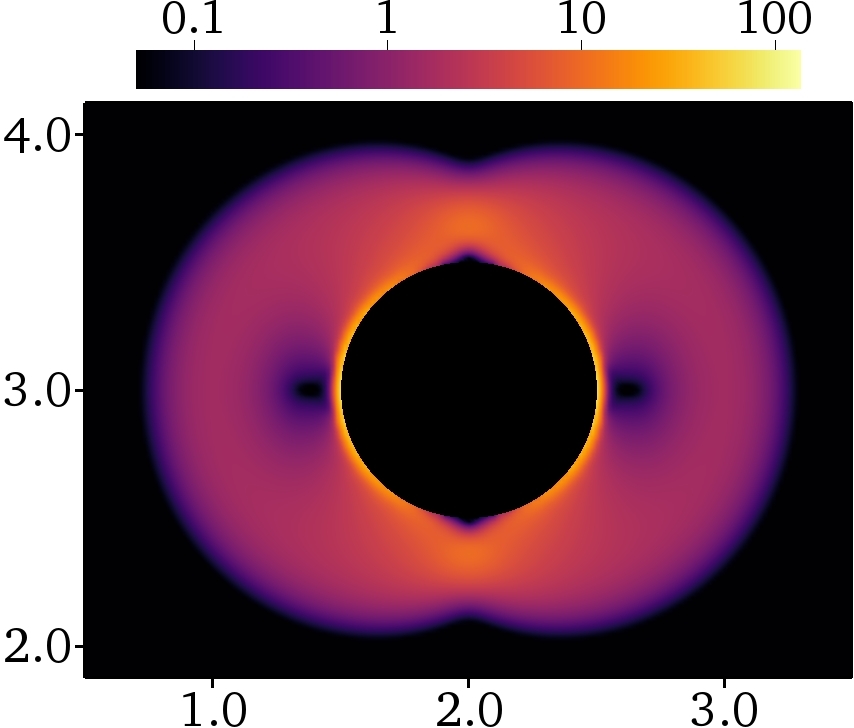}
        \caption
        {
            $\norm{\strainrate}$ (s$^{-1}$)
        }
    \end{subfigure}
    \\[2em]
    \begin{subfigure}{0.49\textwidth}
        \includegraphics[width=\columnwidth]{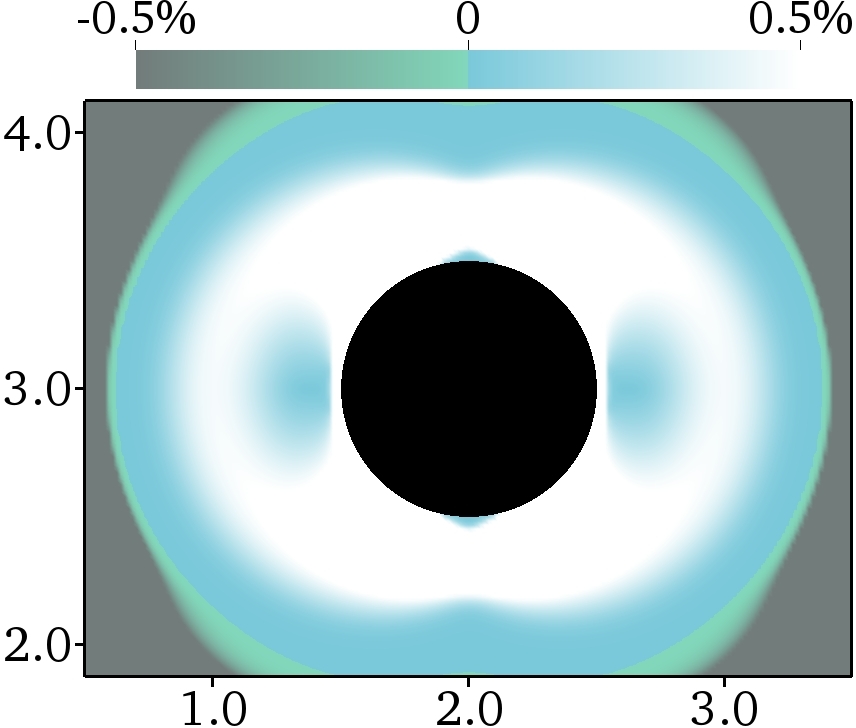}
        \caption
        {
            ${\norm{\stress} / \tau_0 - 1}$
        }
    \end{subfigure}
    \begin{subfigure}{0.49\textwidth}
        \includegraphics[width=\columnwidth]{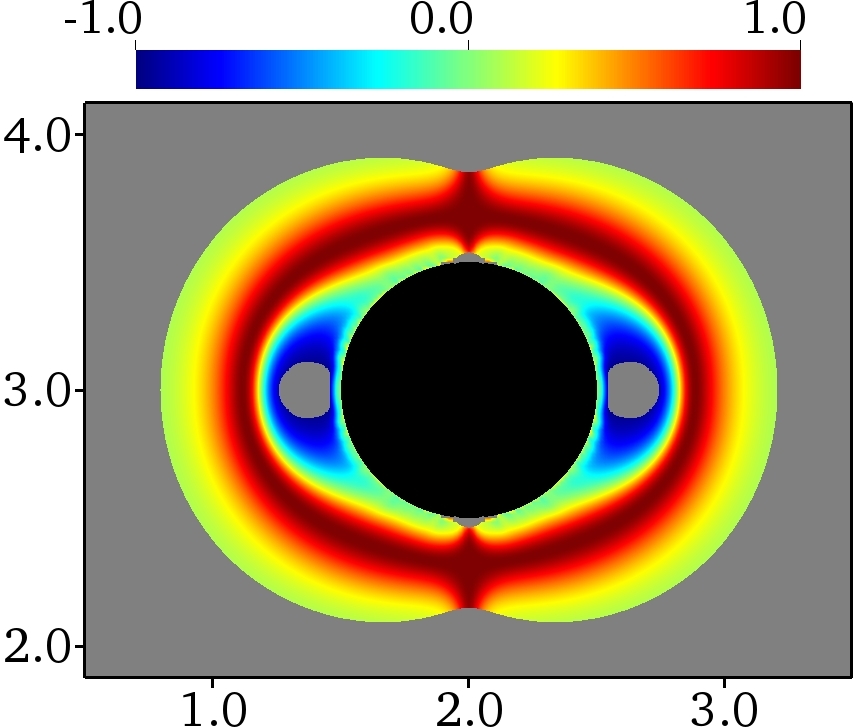}
        \caption
        {
            $\Lambda$
        }
    \end{subfigure}
    \caption
    {
        Steady-state flow field around the sphere (black) in the plane $\hat{y}=2$ for $\Bin =
        340.7$, with computed Stokes drag coefficient $C_S = 387.36$. Depicted are (a) the relative
        velocity field and its magnitude, (b) the rate-of-strain magnitude, (c) the yield surface
        stress deviation and (d) the flow topology parameter. Note that the colormap in (b) has
        logarithmic scaling, and that values outside the colormap for (c) are mapped on to the
        endpoints. In (d), the yield surface computed with $\delta = 10^{-3}$ is also masked out in
        grey. 
    }
    \label{fig:sphere_xzslice}
\end{figure*}

\begin{figure*}
    \centering
    \begin{subfigure}{0.4\textwidth}
        \includegraphics[width=\columnwidth]{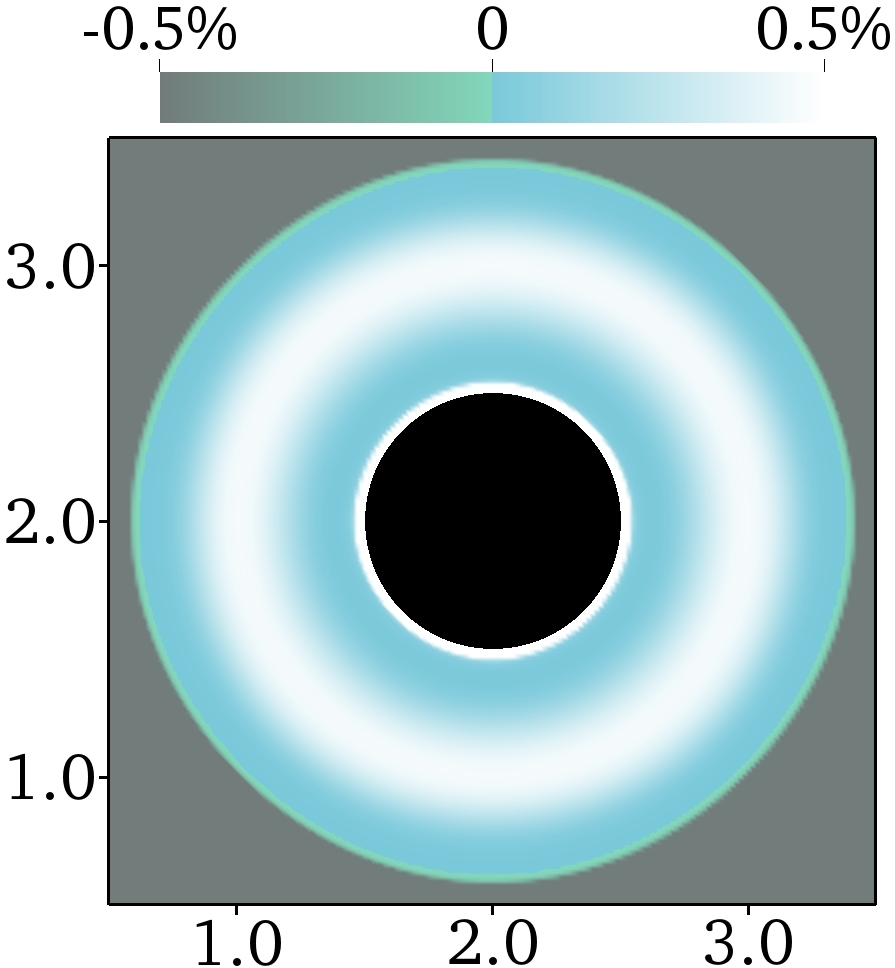}
        \caption
        {
            ${\norm{\stress} / \tau_0 - 1}$
        }
    \end{subfigure}
    \begin{subfigure}{0.4\textwidth}
        \includegraphics[width=\columnwidth]{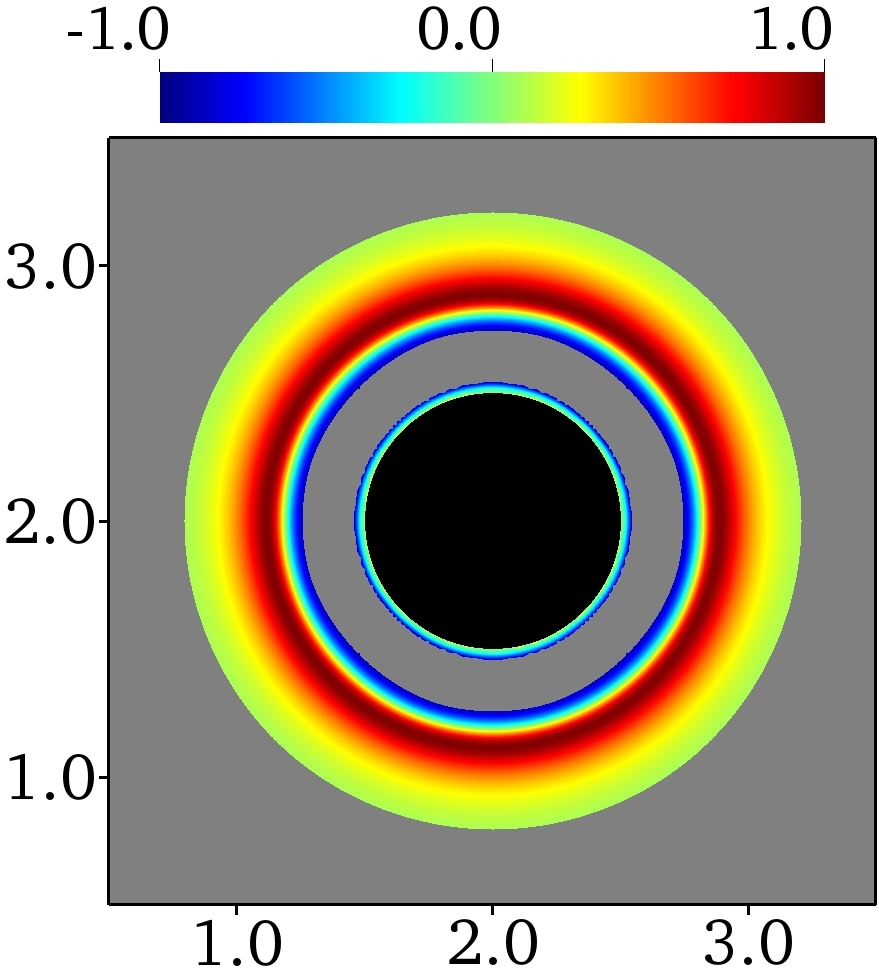}
        \caption
        {
            $\Lambda$
        }
    \end{subfigure}
    \caption
    {
        Steady-state flow field around the sphere (black) in the plane $\hat{z}=2$ for $\Bin =
        340.7$.  Depicted are (a) the yield surface stress deviation and (b) the flow topology
        parameter.  Note that values outside the colormap for (a) are mapped on to the endpoints.
        In (b), the yield surface computed with $\delta = 10^{-3}$ is also masked out in grey.
    }
    \label{fig:sphere_xyslice}
\end{figure*}

In figure \ref{fig:sphere_xzslice}, we show plots of the same variables as those in figure
\ref{fig:cylinder} for the cylinder. There are several noteworthy points to make about the
differences between the two cases. Firstly, the overall size of the yield envelope is significantly
smaller for the sphere than the cylinder, both in the direction of flow and in the $xy$-plane.
Secondly, the unyielded regions within the envelope are much smaller. Both of these observations
are as expected, since the axial symmetry of the sphere causes reduced shearing effects in the plane.
Equivalently, the extended cylinder enhances shearing in the plane, leading to the larger unyielded
regions and genuine rigid body rotation. Finally, the relative stress deviation plot (c) proves
that there aren't any regions in the equatorial torus where the stress is actually below the yield
stress threshold. Although the stress gets extremely close to the yield threshold in these
areas of low strain, there is no evidence of a genuinely unyielded region, since the relative
stress plot stays blue. The diffuse interfaces in this plot shed some light on why the torus-shaped
plug is present in some papers and not others: it is severely close to the threshold value without
actually reaching it, and without a sharp transition near it. Thus, the benefits of the
visualisation strategy introduced by Treskatis becomes clear: the plot in (c) shows in details what
the stress deviation looks like near the yield surface, while the well-known masked out contour
$\norm{\stress} = 1.001 \tau_0$ in (d) oversimplifies the case. This is all in agreement with the
original arguments put forth by Beris et al.~\cite{beris1985creeping} Figure
\ref{fig:sphere_xyslice} shows the last two plots through the slice $\hat{z} = 2$,
i.e.~perpendicular to the flow direction, but still through the centre of the sphere.

\begin{figure*}
    \centering
	\begin{subfigure}{0.3\textwidth}
		\includegraphics[width=\columnwidth]{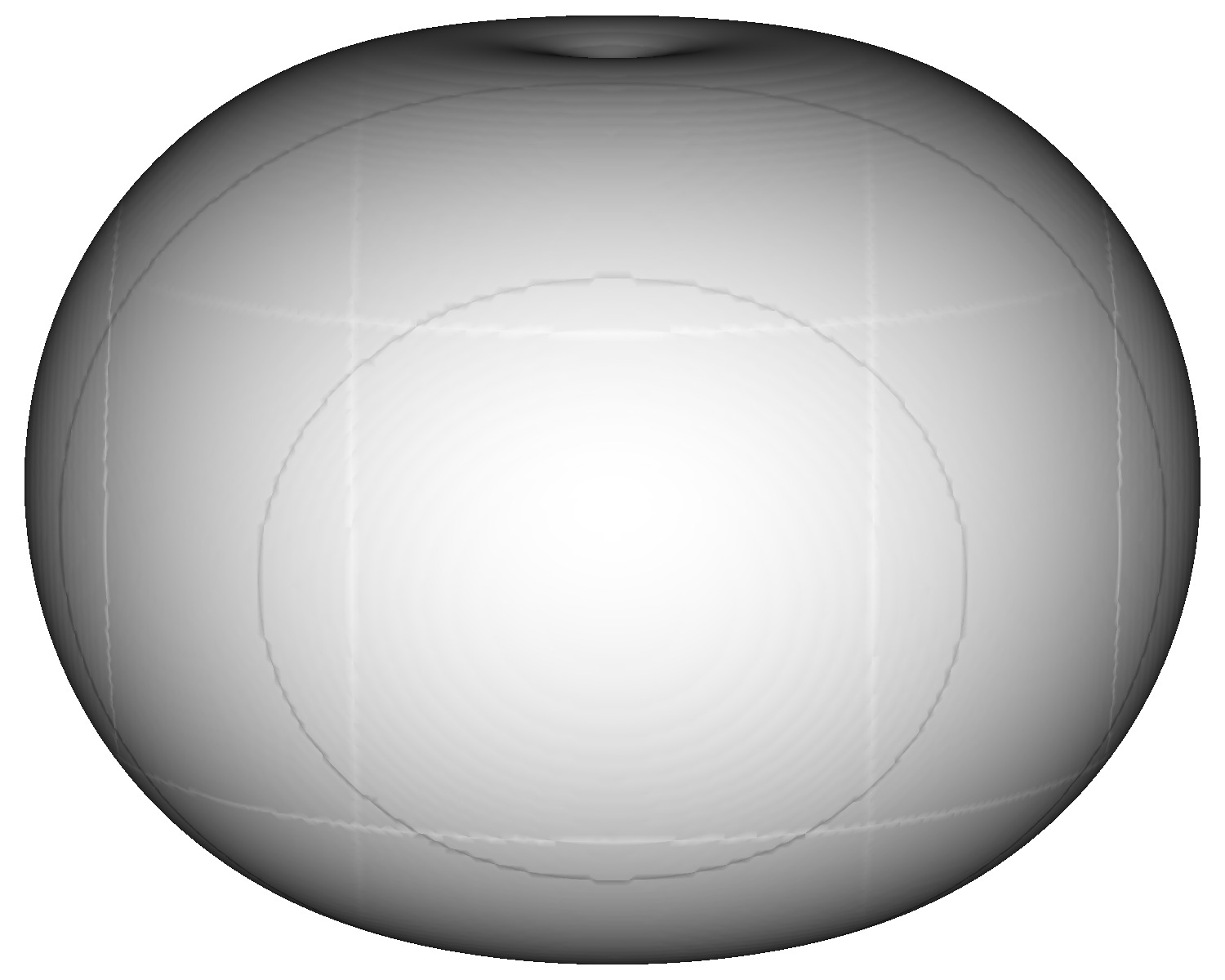}
        \caption{Opacity 100\%}
	\end{subfigure}
	~
	\begin{subfigure}{0.3\textwidth}
		\includegraphics[width=\columnwidth]{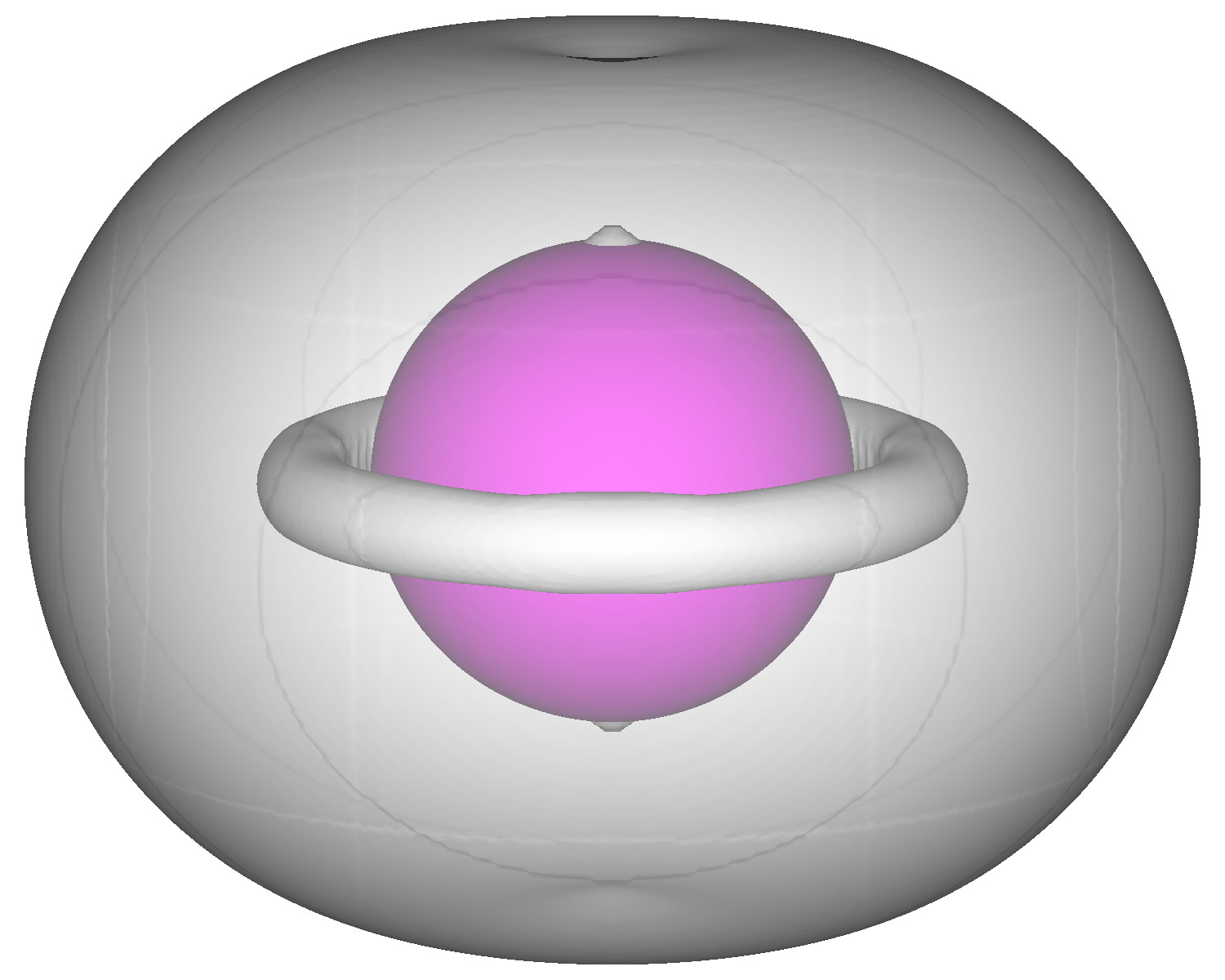}
        \caption{Opacity 50\%}
	\end{subfigure}
	~
	\begin{subfigure}{0.3\textwidth}
		\includegraphics[width=\columnwidth]{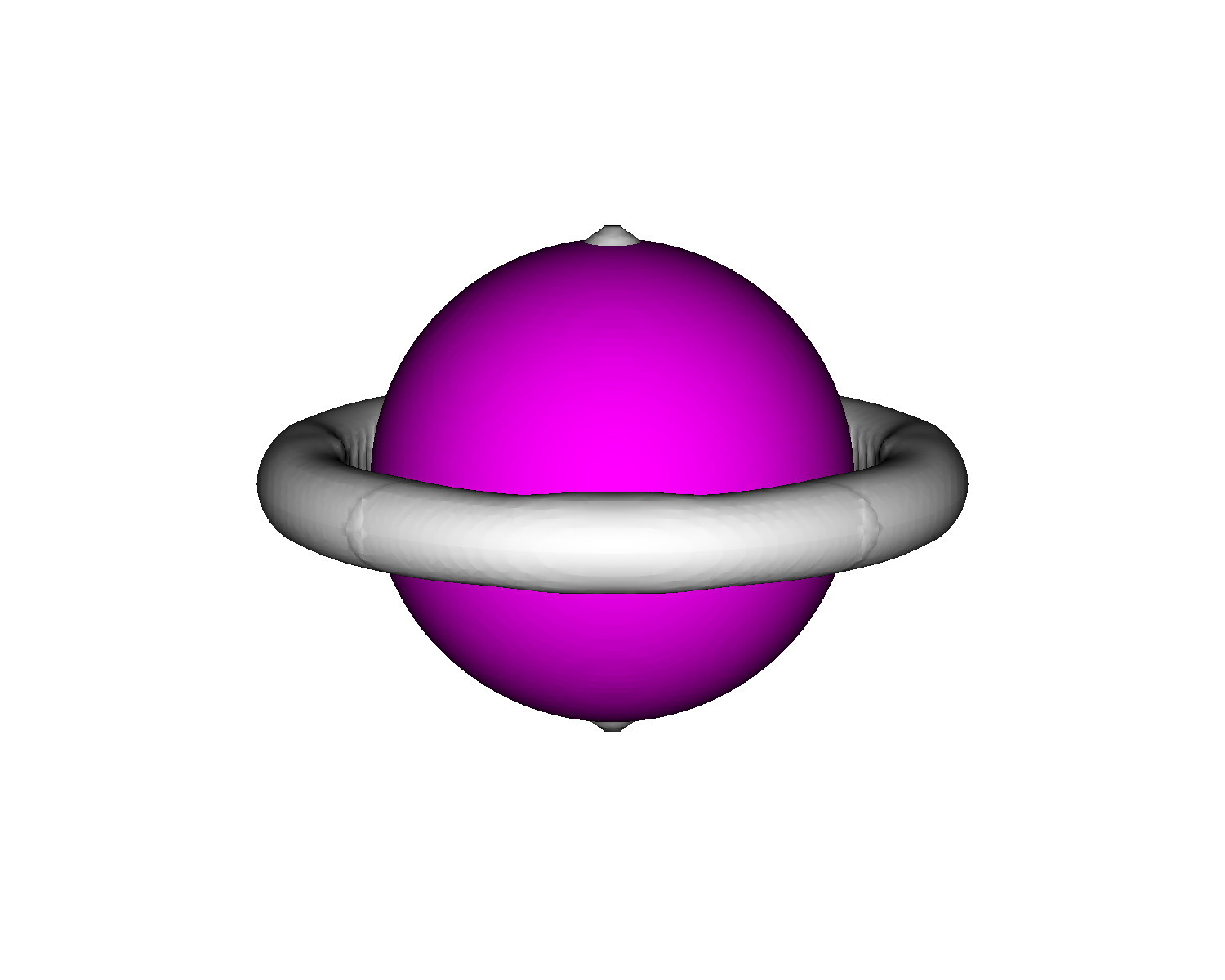}
        \caption{Opacity 0\%}
	\end{subfigure}
	\caption
	{
        Yield surface represented by the contour $\norm{\stress} = 1.001 \tau_0$ (light grey)
        around sphere (magenta) moving at constant speed through a Bingham fluid with $\Bin =
        340.7$. The opacity of the yield envelope is reduced from (a) 100\% to (b) 50\% and finally
        (c) 0\%, in order to reveal the sphere and unyielded regions within it. 	
    }
	\label{fig:sphere_yieldsurface}
\end{figure*}

\begin{figure*}
    \centering
	\begin{subfigure}{0.3\textwidth}
		\includegraphics[width=\columnwidth]{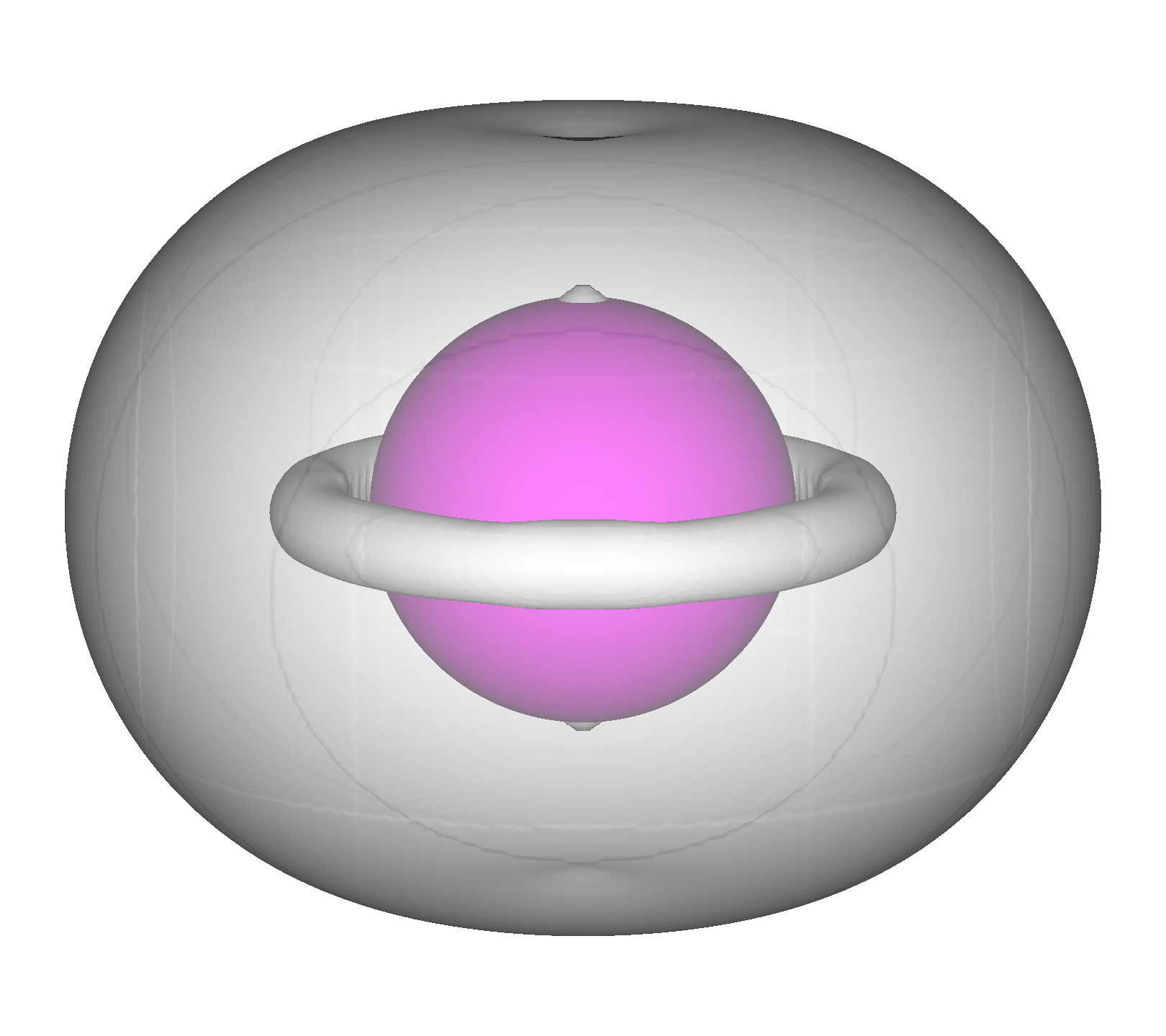}
        \caption{$\delta = 10^{-3}$}
	\end{subfigure}
	~
	\begin{subfigure}{0.3\textwidth}
		\includegraphics[width=\columnwidth]{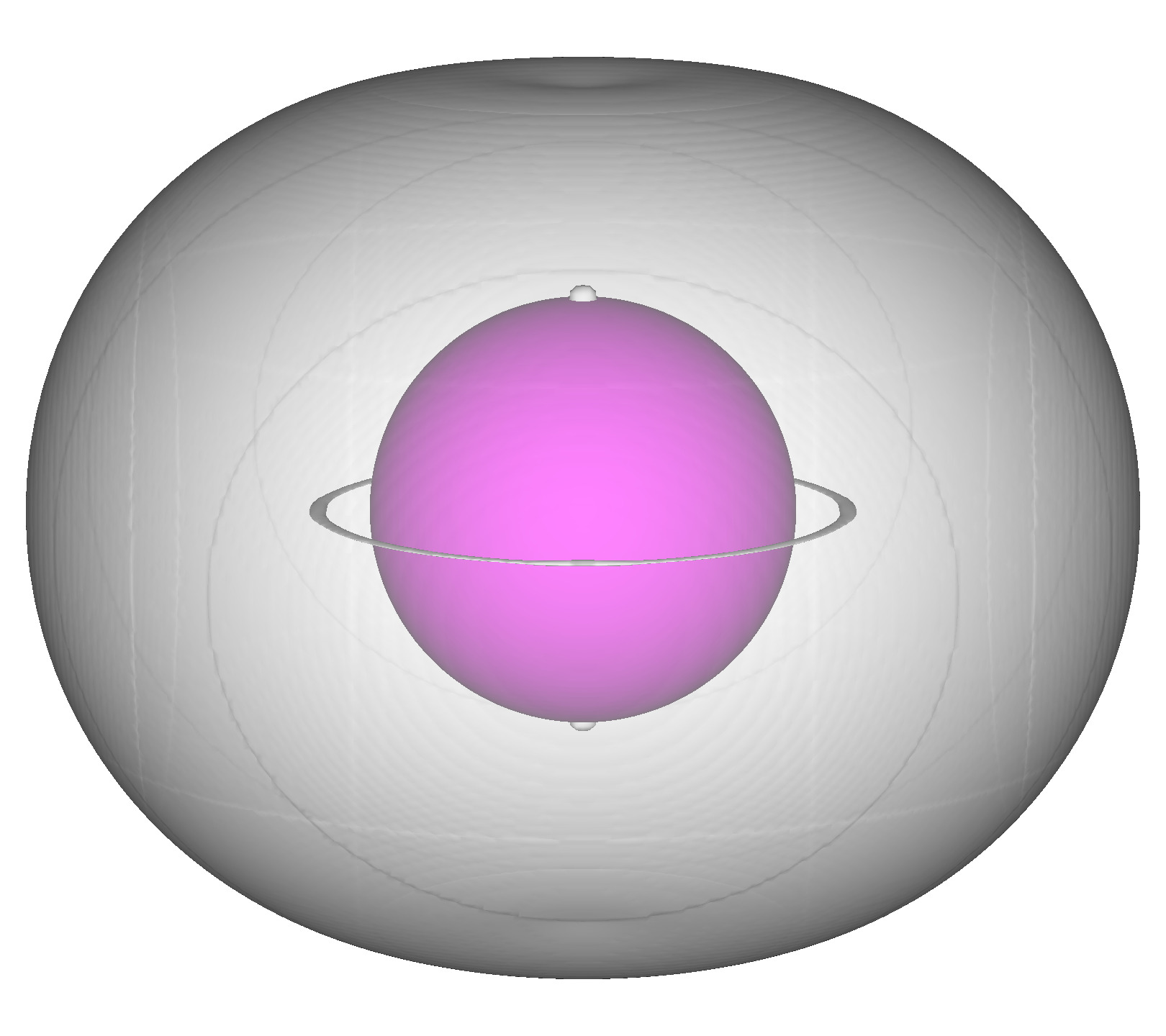}
        \caption{$\delta = 10^{-4}$}
	\end{subfigure}
	~
	\begin{subfigure}{0.3\textwidth}
		\includegraphics[width=\columnwidth]{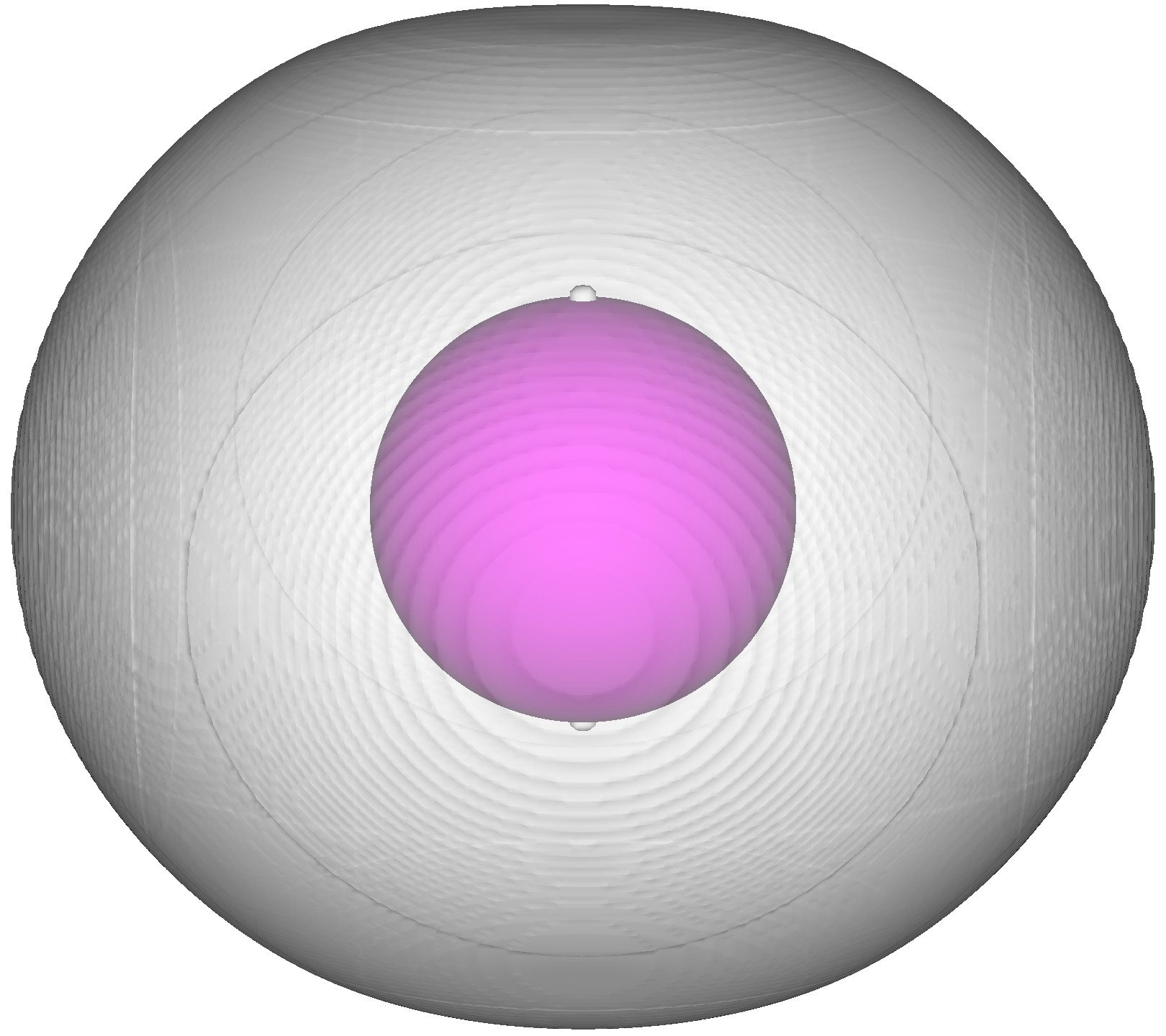}
        \caption{$\delta = 0$}
	\end{subfigure}
	\caption
	{
        Yield surface represented by the contour $\norm{\stress} = (1 + \delta) \tau_0$ (light
        grey) around sphere (magenta) moving at constant speed through a Bingham fluid with $\Bin =
        340.7$. The effect of varying $\delta$ is evident, as it is reduced from (a) $10^{-3}$ to
        (b) $10^{-4}$ and finally (c) $0$. 
    }
	\label{fig:sphere_yieldsurface_delta}
\end{figure*}

Finally, in figures \ref{fig:sphere_yieldsurface} and \ref{fig:sphere_yieldsurface_delta}, we
visualise three-dimensional stress contours in the vicinity of the sphere. In the former, the yield
surfaces are all computed with $\delta = 10^{-3}$. Since the yield envelope fully encloses the
sphere and plug regions, its opacity is reduced from subfigures (a) to (c), in order to reveal the
shape of the enclosed yield surfaces.  This kind of visualisation is entirely novel to the best of
our knowledge, and gives a richer picture of the yield surface topology. Since there is no
body-fitted mesh or assumption on symmetry in the flow, this type of simulation opens a range of
possibilities for investigating flow patterns and yield surfaces in complex configurations. The
effect of reducing $\delta$ is illustrated in figure \ref{fig:sphere_yieldsurface_delta}, and
confirms that the toroidal yield surface at the equator disappears in the limit $\delta \rightarrow
0$, at the cost of poorer resolution for the yield envelope. 

\subsection{Non-trivial particle shape}

\begin{figure*}
    \centering
	\begin{subfigure}{0.45\textwidth}
		\includegraphics[width=\columnwidth]{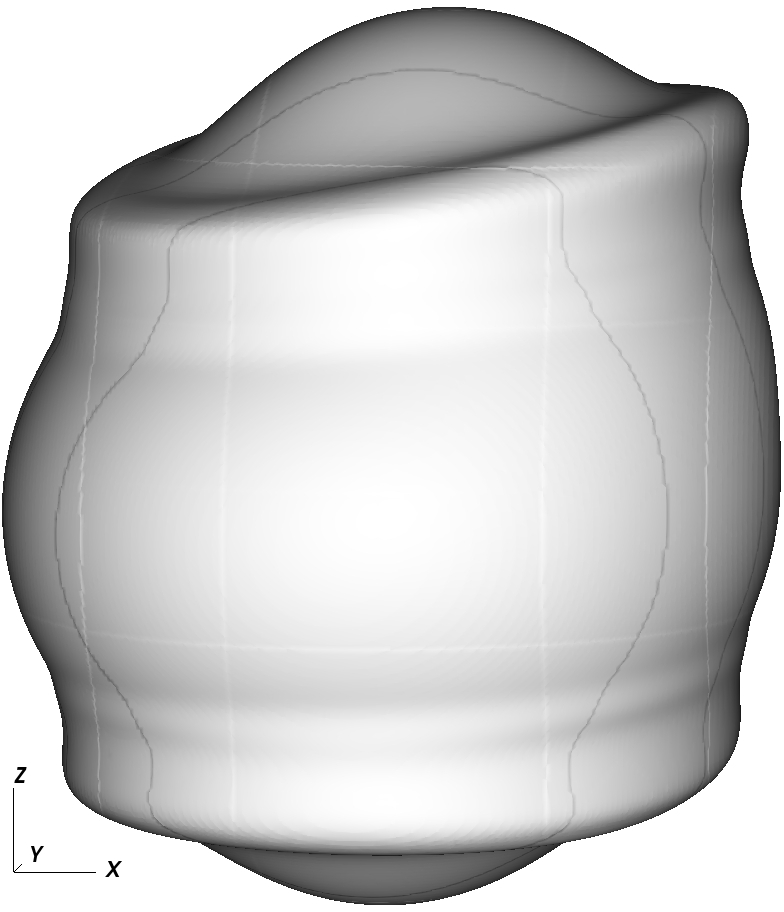}
        \caption{Opacity 100\%}
	\end{subfigure}
	~
	\begin{subfigure}{0.45\textwidth}
		\includegraphics[width=\columnwidth]{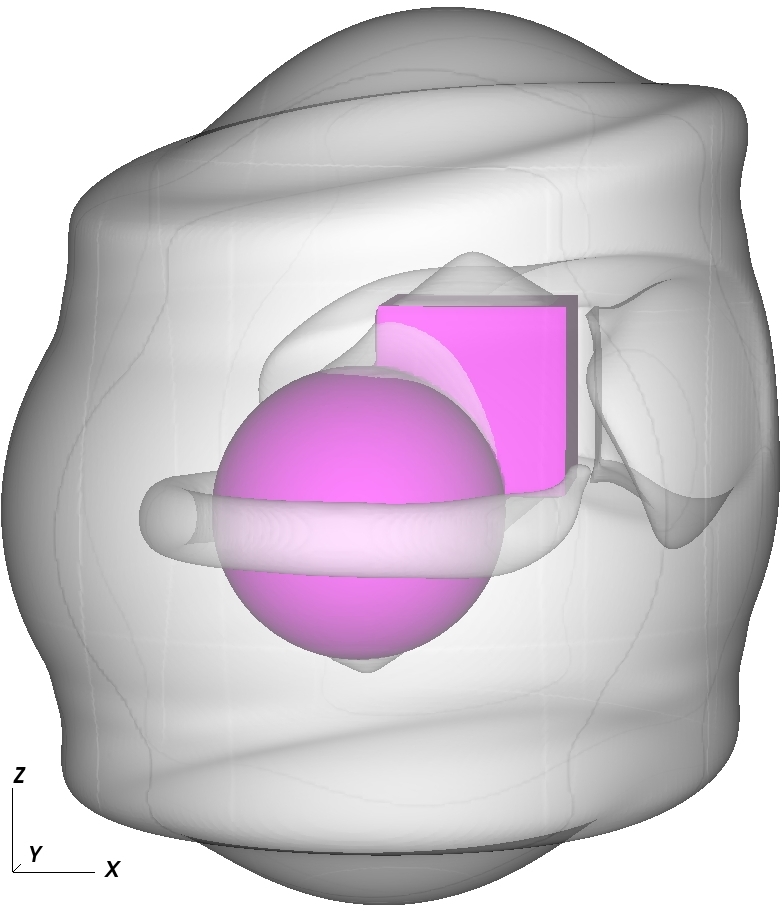}
        \caption{Opacity 50\%}
	\end{subfigure}
	\caption
	{
        Yield surface computed with $\delta = 10^{-3}$ for the flow of a Bingham fluid around the
        union of a sphere and a cube with $\Bin = 340.7$. The opacities of the stress contours are
        (a) 100\% and (b) 50\%, in order to show unyielded regions enclosed by the outer yield
        envelope. (Multimedia view)
	}
	\label{fig:spherecube}
\end{figure*}

As a final demonstration, we replace the sphere by an object which is the union of a sphere and
cube, and orient it so that the flow field becomes asymmetrical. This means that a
three-dimensional representation is necessary to capture the fluid dynamics. To be precise, the
sphere is the same size as in the previous section, but has centre coordinates $(\hat{x}, \hat{y},
\hat{z}) = (1.8, 1.8, 2.8)$, while the cube is defined by two points on opposing sides of its main
diagonal, in dimensionless units $(1.85, 1.85, 2.85)$ and $(2.5, 2.5, 3.5)$. We retain the highest
resolution settings used for the sphere, i.e.~$\Omega = [0,4 \mathcal{D}]^2 \times [0, 6
\mathcal{D}]$, $\Delta x = \mathcal{D} / 64$ and $\varepsilon = 10^{-3}$. The resulting 3D yield surface,
computed with $\delta = 10^{-3}$, is shown in figure \ref{fig:spherecube} (Multimedia view). In
contrast to figure \ref{fig:sphere_yieldsurface}, we reduce the opacity of all the yield surfaces,
and not just the enclosing envelope. This is due to the difficulty in separating the locations of
the yield surface types by simple rules. Although the stress contours are now much more complex, we
can still recognise the expected traits: an enclosing yield envelope surrounding the entire body,
in addition to smaller unyielded plugs attached to it at places of low strain rate. In particular,
there are caps of unyielded material fore and aft of the object in the flow direction, as well as
along the narrow intersection of the sphere and cube. Additionally, the characteristic low-strain
rotating region around the sphere's equator is visible, but it does not continue around the entire
object.  The free cube sides, which are aligned with the flow direction, only lead to a very narrow
boundary layer separating the object from the yield envelope. A video is available online, in which
the observer's point of view is rotated around the object in order to show the yield surface from
all polar angles. 

\section{Conclusions}
\label{sec:conclusions}

We have presented a new methodology for simulating highly resolved flow of generalised Newtonian
fluids (in particular, viscoplastics) in three-dimensional domains with non-rectangular boundaries.
The domain geometry is treated using embedded boundaries, which are applied to viscoplastic fluid
computations for the first time. Highly resolved simulations are made possible by implementing the
algorithm in AMReX, a high performance library of tools for solving PDEs on structured grids. The
solver is robust and efficient, and allowed us to perform simulations of various objects moving at
constant relative speed through an infinite volume of Bingham fluid, revealing the shape of the
resulting yield envelopes and plug regions. We hope our contribution can prove a useful tool to
researchers interested in studying three-dimensional flows of Bingham fluids in non-rectangular
domains or around various body shapes.



\begin{acknowledgments}
K.~S.~would like to acknowledge the EPSRC Centre for Doctoral Training in Computational Methods for
Materials Science for funding under grant number EP/L015552/1. Additionally, he acknowledges the
funding and technical support from BP through the BP International Centre for Advanced Materials
(BP-ICAM) which made this research possible. Finally, he is grateful to Arndt Ryo Koblitz for
useful discussions and for sharing his valuable insights.
\end{acknowledgments}

\appendix

\section{Polynomial interpolation}

Given a set of $n+1$ real values $y_i$ at data points $x_i$ for $i \in \{0,\dots,n\}$, polynomial
interpolation seeks to find a polynomial function of degree $n$ whose curve passes through each of
them. The Lagrange form of the interpolant is given by
\begin{equation}
    f(x) = \sum_{i=0}^n y_i \prod_{j \neq i} \frac{x-x_j}{x_i-x_j} ,
    \label{eq:lagrange_polynomial}
\end{equation}
so that it's derivative is
\begin{equation}
    \frac{\partial f}{\partial x}
    = \sum_{i=0}^{n} y_i \frac{\sum_{j \neq i}(x-x_j)}{\prod_{j \neq i}(x_i - x_j)}
    = \sum_{i=0}^{n} y_i \frac{n x - \sum_{j \neq i} x_j}{\prod_{j \neq i}(x_i - x_j)} .
    \label{eq:lagrange_polynomial_gradient}
\end{equation}
The polynomial interpolant is unique and accurate to order $n$.

\section{Analytical solution for Poiseuille flow of a Papanastasiou regularised Bingham fluid}
\label{app:papabing}

Plane Poiseuille flow of a Papanastasiou regularised Bingham fluid in a channel of width $2
\mathcal{W}$ in the $z$-direction is driven by a constant applied pressure gradient $\mathcal{G}$
in the $x$-direction, resulting in a steady-state solution with non-zero $x$-velocity aligned with
the channel plates. The velocity profile is only a function of $z$, and takes its maximum value at
the centre of the channel, $z=0$. The plane $z = 0$ is a symmetry plane for the solution, so we
solve for $z \geq 0$. We take $\mathcal{W}$ as the characteristic length, while the
characteristic speed $\mathcal{U}$ is the maximum of the unregularised analytical solution. Scaling
by these, in addition to strain rate $\mathcal{U} / \mathcal{W}$ and stress $\mu \mathcal{U} /
\mathcal{W}$, the steady-state governing equations in dimensionless form are 
\begin{subequations}
\begin{align}
    \label{eq:papabing:governing}
    \nabla p &= \nabla \cdot \stress , \\
    p &= - \frac{\mathcal{G} \mathcal{W}^2}{\mu \mathcal{U}} x , \\ 
    \stress &= \left( 1 + \Bin \, 
    \frac{1 - \exp \left( -\frac{\norm{\strainrate}}{ \varepsilon} \right)}{\norm{\strainrate}}
    \right)
    \strainrate .
\end{align}
\end{subequations}
Here, the Bingham number is $\Bin = \tau_0 \mathcal{W}/(\mu \mathcal{U})$.  Since the only non-zero
velocity component is $u = u(z)$, the magnitude of the rate-of-strain tensor is $\norm{\strainrate}
= |\partial u / \partial z|$. Furthermore, since $u$ is monotonically decreasing for $z > 0$, we
have $|\partial u / \partial z| = -\partial u / \partial z$. With this information in place, the
first component of \eqref{eq:papabing:governing} is 
\begin{equation}
    - \frac{\Bin}{z_0} 
    = \frac{\partial}{\partial z} 
    \left( \frac{\partial u}{\partial z} 
    \left( 1 - \Bin \, 
        \frac{1 - \exp \left( \frac{\partial u / \partial z}{\varepsilon} \right)}
        {\partial u / \partial z} \right)
    \right) ,
    \label{eq:papabing:first-component}
\end{equation}
where $z_0 = \tau_0 / ( \mathcal{G} \mathcal{W} )$ is the location of the yield surface in the
unregularised Bingham model. Its relationship to the Bingham number (which is superfluous in this
description) is 
\begin{equation}
    Bi = \frac{2 z_0}{(1 - z_0)^2} .
\end{equation}
Equation \eqref{eq:papabing:first-component} is a separable differential equation for $\partial u /
\partial z$:
\begin{equation}
    - \frac{2}{(1 - z_0)^2} 
    = \frac{\partial^2 u}{\partial z^2} 
    \left( 1 + \frac{2 z_0}{\varepsilon (1 - z_0)^2} \exp \left(
        \frac{\partial u / \partial z}{\varepsilon} \right) \right) ,
    \label{eq:papabing:separable}
\end{equation}
For ease of writing, we introduce $\xi = \frac{2 z_0}{\varepsilon (1 - z_0)^2}$.
The solution to \eqref{eq:papabing:separable} is then 
\begin{equation}
    \frac{\partial u}{\partial z} 
    = - \frac{2 (z - z_0)}{(1 - z_0)^2} 
    - \varepsilon W \left( \xi e^{-\xi (z / z_0 - 1)} \right) ,
    \label{eq:papabing:uy}
\end{equation}
where we have found the explicit solution by use of the Lambert $W$ function and the fact that
$\partial u / \partial z = 0$ at $z=0$. Using the no-slip boundary condition $u(1) = 0$, and the
identity 
\begin{equation}
    \int W \left( A e^{Bx} \right) {\rm d} x = \frac{1}{2B} \left( 1 + W \left( A e^{Bx} \right)
    \right)^2 + C ,
\end{equation}
direct integration then gives the analytical expression for the velocity profile as
\begin{widetext}
\begin{equation}
    u (z) = 1 - \left( \frac{z - z_0}{1 - z_0} \right)^2    
    + \frac{z_0 \varepsilon}{2 \xi}
    \left( 
    \left( 1 + W \left( \xi e^{-\xi (z/z_0 - 1)} \right) \right)^2 
    - \left( 1 + W \left( \xi e^{-\xi (1/z_0 - 1)} \right) \right)^2 
    \right) .
\end{equation}
\end{widetext}

\section{Faxén's formula}

As shown by Fax{\'en} in 1946 \cite{faxen1946forces}, the creeping drag force per unit length of a
cylinder with diameter $\mathcal{D}$ moving at speed $\mathcal{U}$ through the centre of a channel
of width $\mathcal{W}$ filled with a Newtonian fluid with viscosity coefficient $\mu$, is given by  
\begin{widetext}
\begin{equation}
    F_D = \frac{4 \pi \mu \mathcal{U}}{\ln \WD - 0.9157 + 1.7244 \DW^2 - 1.7302 \DW^4 + 2.4056
    \DW^6 - 4.5913 \DW^8} .
    \label{eq:faxen}
\end{equation}
\end{widetext}

\bibliography{References.bib}

\end{document}